\documentclass[trackchanges, manuscript]{aastex701}

\usepackage{newtxtext,newtxmath}
\usepackage{graphicx}	% Including figure files
\usepackage{amsmath}	% Advanced maths commands
\usepackage{tikz}
\usetikzlibrary{positioning, shapes.geometric, arrows}
\usetikzlibrary{arrows.meta, positioning}
\usepackage{footmisc}
\usepackage{adjustbox}
\usepackage{array}
\usepackage{caption}
\usepackage{booktabs}
\usepackage{xcolor}
\usepackage{colortbl}
\usepackage{fixltx2e}
\usepackage[T1]{fontenc}
\usepackage{url}
\usepackage[utf8]{inputenc}  % (for pdflatex)
\usepackage{hyperref}

\begin{document}

\title{A Convolutional Neural Network-Derived Catalog of Solar Flares from Soft X-Ray Observations}

\author[orcid=0000-0002-6008-2277,sname='NF']{Nastaran Farhang}
\affiliation{Sydney Institute for Astronomy, University of Sydney, NSW 2006, Australia}
\email[show]{nastaran.farhang@sydney.edu.au}  

\author[orcid=0000-0001-5100-2354,gname=Bosque, sname='MW']{Michael. S. Wheatland} 
\affiliation{Sydney Institute for Astronomy, University of Sydney, NSW 2006, Australia}
\email[]{michael.wheatland@sydney.edu.au}

\author[orcid=0000-0003-4642-141X, gname=Savannah,sname='AM']{Andrew Melatos}
\affiliation{School of Physics, University of Melbourne, Parkville, Victoria, 3010, Australia}
\affiliation{OzGrav-Melbourne, Australian Research Council Centre of Excellence for Gravitational Wave Discovery,
Parkville, Victoria, 3010, Australia}
\email[]{amelatos@unimelb.edu.au}

\collaboration{3}{}

\begin{abstract}
A convolutional neural network (CNN) is used to construct a new catalog for solar flares based on high resolution (1-s cadence) Geostationary Operational Environmental Satellites (GOES) soft X-ray data. The CNN is trained to identify flare rise episodes. From 1 January 2018 to 22 August 2025, the algorithm detects 111,580 flare candidates, compared with 14,612 events in the corresponding GOES catalog. For each candidate, the probability of being a true positive is quantified by Bayesian inference based on the peak flux, rise time, and temporal coincidence with cataloged events where available. The flare size and waiting-time distributions are studied and compared with the GOES catalog. The CNN catalog shows a steeper power-law index for raw peak fluxes ($-2.59 \pm 0.02$) than GOES ($-2.25 \pm 0.04$), indicating the CNN’s higher sensitivity to small events. After background correction, the indices are $-1.97 \pm 0.02$ (CNN) and $-2.05 \pm 0.04$ (GOES). The CNN catalog extends the power-law distribution of flare peak fluxes by one order of magnitude at the small-flux end compared with the GOES background-subtracted catalog. A Bayesian blocks analysis of the waiting-time distributions from the GOES and CNN catalogs indicates broad consistency with a piecewise Poisson process. We find that previously reported correlations between flare sizes and waiting times are significantly influenced by obscuration, that is, under-counting weaker or overlapping flares during periods of elevated flux. The new CNN catalog provides a foundation for complete and consistent studies of solar flare statistics.

\end{abstract}


\section{Introduction}

Solar flares are sudden releases of energy in the Sun's atmosphere, that provide emissions across a range of wavelengths arising from distinct physical processes \citep{fletcher2011observational, benz2017flare, fletcher2024solar}. Despite decades of investigation, the trigger mechanism of solar flares remains uncertain. Flare statistics, especially the relation between event sizes (generally a measure of flare magnitude, here defined as the peak soft X-ray (SXR) flux) and waiting times within individual active regions (ARs), are important for understanding flare energy storage and release \citep{mike2001, wheatland2009monte, kanazir2010time, Lippiello2010, biasiotti2025statistical}. 

In practice, studies of flare statistics sometimes disagree about the distributions of sizes, waiting times, and rise times. For instance, \cite{mike2001, Moon2001} and \cite{wheatland2008energetics} characterized flare waiting times by nonstationary Poisson models and found no robust correlation between flare sizes and waiting times. Other analyses identified deviations from Poisson behavior, evidenced by periods of elevated event rates and correlations between sizes and waiting times of events \citep{lepreti2001solar, Lippiello2010, Hudson2020}. \cite{boffetta1999power} and \cite{aschwanden2010reconciliation} argued that aggregating ARs across the solar disk over longer timescales yields a waiting-time distribution with a power-law tail. Self-organized critical models reproduce avalanches of magnetic reconnection events with power-law size and Poisson waiting-time distributions, but avalanche models are not physical \citep{lu1991avalanches, farhang2018}. \cite{verbeeck2019solar} argued that background-subtracted peak fluxes are better described by a lognormal distribution than a power law. Monte-Carlo models, which simulate flares as a random process driven by gradual energy accumulation in ARs, best match observations when configured with no dependence between the event rate and stored energy, yielding Poisson-like waiting times \citep{wheatland2009monte, kanazir2010time}. Stress-relax models \citep{fulgenzi2017radio, Carlin2023} suggest that a flare reduces the system’s available free energy, i.e., the stored magnetic energy available for release, with larger events causing greater depletion and lengthening the reaccumulation time. These models predict a correlation between flare sizes and waiting times. It remains uncertain whether flare statistics reflect a spatially uncorrelated process, such as self-organized criticality, or result from correlated stress relaxation cycles within ARs.

The lack of consensus in observational studies may be due to many factors, including the sensitivity of flare statistics to the observing band, the period of observation and its relationship to the phase of the solar cycle, the spatial scope of the analysis (individual ARs versus the full solar disk), and the flare catalog employed. Different flare catalogs are constructed using different flare detection algorithms (FDAs), which incorporate various assumptions and are tuned to different observing bands. Consequently, catalogs vary substantially in the number, timing, and intensity of events (see Appendix \ref{app3} for a summary of the most widely used catalogs and their FDAs).

The most commonly used flare catalog is the National Oceanic and Atmospheric Administration (NOAA) Geostationary Operational Environmental Satellites (GOES) flare catalog \citep{GOES_R_XRS_L2_Products_Guide}, which is constructed based on one-minute averaged SXR fluxes in the 1–8\AA~ channel. The FDA for the GOES catalog defines an event to start when four consecutive one-minute flux values exceed a background threshold and rise monotonically with the last exceeding the first by $\geq 40\%$. The peak of the flare is the flux maximum, and the end is marked when the flux declines to half the peak value \citep{swalwell2018reported}. 

The criteria used to construct the GOES catalog introduce systematic bias. In particular, during phases of high solar activity, the enhanced background flux suppresses the detection of weaker events and masks temporally adjacent flares, a limitation termed obscuration \citep{mike2001, mike2002}. As a result, flare occurrence rates and associated statistics derived from the GOES catalog are prone to underestimation. Furthermore, the start and end times of events are marked according to arbitrary rules, and the registered peak fluxes incorporate the total SXR flux from all ARs on the solar disk, which leads to inflated flux values during solar maximum, when multiple ARs exist at any moment.

In this study we address obscuration by introducing a new FDA based on convolutional neural networks (CNNs) to identify flaring episodes directly from full-resolution GOES SXR time series (rather than one-minute averaged data). CNNs offer a promising alternative to traditional FDAs by automatically learning multiscale features directly from raw data through multiple hierarchical layers \citep{farabet2012learning, zeiler2014visualizing}. They capture global and local patterns, and scale to large datasets and high-dimensional inputs \citep{ismail2019deep, yang2025}. CNNs have been used recently to identify flare precursors and study magnetic field evolution from solar magnetograms, extreme ultraviolet (EUV) images, and SXR data \citep{park2018, li2020, bhattacharjee2020supervised, abed2021automated, zheng2021hybrid, li2022knowledge, landa2022low, sun2022, wei2023prediction, sun2023deep, li2023deep, hassani2025solar}. However, their application to constructing a flare catalog from SXR data remains unexplored. To our knowledge, this paper is the first study to develop a CNN for this purpose.

The CNN yields a catalog containing more than seven times as many events as the GOES catalog. Using this catalog, we revisit the statistical properties of the identified events and contrast them with earlier reports. The paper is structured as follows. Section \ref{sec:obs} describes the data used to construct the catalog. Section \ref{sec:cnn} outlines the CNN framework, including the reference catalog (Section \ref{sec:refcat}), training and validation (Section \ref{sec:transet}), data analysis workflow as applied to the full dataset (Section \ref{sec:testset}), and the Bayesian approach used for confidence estimation (Section \ref{sec:bayesian}). Section \ref{sec:res} presents the CNN catalog, quoting the detection confidence (Section \ref{res:conf}), the statistical properties of peak fluxes over a solar cycle (Section \ref{sec:stat1} and \ref{sec:stat2}), and the size and waiting-time distributions of events (Sections \ref{sec:stat3}). Section \ref{sec:obscur} quantifies obscuration and demonstrates that the earlier reports of correlations between flare sizes and waiting times are biased. Finally, Section \ref{sec:conclusion} presents conclusions.

\section{Data} \label{sec:obs}
The multi-satellite GOES mission collects SXR real-time data for space weather forecasting \footnote{\url{https://www.goes-r.gov/downloads/resources/documents/GOES-RSeriesDataBook.pdf}}. The GOES mission has been continuously monitoring the Sun in SXR for over five decades, with 19 satellites launched to date. To ensure continual data coverage, at least two operational GOES satellites are always in position serving as the primary observer and the backup. This dual-satellite configuration minimizes data loss due to interruptions such as the daily Earth occultation of each satellite’s line of sight to the Sun, which can last up to approximately 70 minutes. When one satellite experiences a coverage gap, measurements from the other instrument, provided they are flagged as``good quality'' data, are incorporated to maintain a continuous dataset.

The GOES XR Sensors (XRS) measure solar X-ray flux in two channels: the short channel (XRS-A), which covers 0.5 to 4 \AA~ with a sensitivity range of \(10^{-9}\) to \(10^{-3}\) Wm\(^{-2}\) (3 to 25 keV), and the long channel (XRS-B), spanning 1 to 8 \AA~ with a sensitivity range of \(10^{-8}\) to \(10^{-3}\) Wm\(^{-2}\) (1.5 to 12.4 keV). Each channel has low irradiance sensors (XRS-A1 and XRS-B1) for measuring low SXR fluxes, and high irradiance sensors (XRS-A2 and XRS-B2) for measuring high SXR fluxes and approximately locating flares on the solar disk. At any moment, one of the A sensors and one of the B sensors are chosen as the main channels based on the measured flux level and a switching threshold. Data from both channels are available at a temporal resolution of one second and as one-minute averages.

Generally, two types of GOES data are available: operational products used by the Space Weather Prediction Center (SWPC) and science-quality data. According to the NOAA guidelines, science-quality data should be used when available. This preference is due to the inclusion of recovered missing data in SWPC real-time operational products, accurate time stamps, high-time-resolution data, one-minute and daily average irradiances, quality flags, flare lists, event summaries, flare location information, and corrections for previously applied SWPC scaling factors. Science-quality data is available for the most recent satellites since 2018. The specific caveats applying to the data are detailed in \cite{GOES8_15_XRS_Readme, Hudson_2024}, and \cite{ Janssens2025}.

In this study, we examine the period from 1 January 2018 to 22 August 2025, during which time the GOES-R series satellites were operational, providing data in units of Wm\(^{-2}\). On the eastern side (Atlantic), GOES-16 served as the primary satellite until its replacement by GOES-19 in April 2025. On the western side (Pacific), GOES-17 was the primary satellite from February 2019 until January 2023, after which GOES-18 assumed that role.

\section{Identifying flares with a CNN} \label{sec:cnn}
In this section, we present a CNN-based framework for identifying solar flares directly from high-resolution GOES SXR data. The framework is designed to automatically detect flare rise episodes. The following subsections detail the construction of the reference catalog used for training and evaluation (Section \ref{sec:refcat}), the training and validation sets (Section \ref{sec:transet}), the procedures for training, validation, and testing (Section \ref{sec:testset}), and the method by which probabilities are assigned to each identified (Section \ref{sec:bayesian}). The CNN architecture is described in Appendix~\ref{app4}.

\subsection{A Reference Catalog for Supervised Flare Identification} \label{sec:refcat}
Supervised learning relies on a well-defined reference database that provides labels with uniform criteria across all samples \citep{zhu2005semi, haeusser2017learning}. In the context of flare detection, the principal identifying features include the start, peak, and end times of events, the morphology of the flux profile during flares, and the characterization of the background level to distinguish quiescent from flaring phases. In practice, however, the construction of such a database presents significant challenges. Solar SXR emissions vary in complex ways with time, and no clear-cut definition of the background flux exists, which unambiguously distinguishes flares from the quiescent level \citep{veronig2004solar, Sadykov_2019, adithya2021solar}. Furthermore, flare lightcurves differ substantially in their temporal profiles \citep{Benz2008, gryciuk2017flare, reep2023understanding}. Several flare catalogs provide start, peak, and end times; however, significant discrepancies exist among them owing to the arbitrary criteria in each FDA (see Appendix \ref{app3} for more details).

The use of an incomplete catalog as a training reference risks introducing systematic bias into the CNN identifications \citep{blanzeisky2021algorithmic}. To mitigate this risk, a new, internally consistent reference catalog is constructed. A total of 145 days are randomly selected from the interval 1 January 2018 to 31 May 2025, spanning solar minimum to maximum. Each day's data are visually inspected, and flare rise episodes are manually identified. 

To ensure coherent and uniform application of the selection criteria, all visual inspections and event identifications are performed by a single analyst (NF). Multiple annotators can introduce inconsistencies arising from differing subjective thresholds and interpretations. On the other hand, although a single-analyst approach minimizes such inconsistencies, it may introduce biases associated with subjective judgment or human error. To address this, an independent cross-check was conducted by a second analyst (MW), who inspected blindly a representative subset of the time interval covered by the reference catalog. The independent inspection broadly confirmed the events identified in the reference catalog, while including a number of additional weak events (with mean background-subtracted peak fluxes below $3 \times 10^{-7}$ Wm$^{-2}$, which NF excluded deliberately). Over-emphasizing such weak events in the training set causes the CNN to confuse random noise or quasi-periodic pulsations with genuine events, leading to spurious transitions between states. In this context, we consider the manual selection performed by NF to provide a sufficiently balanced and appropriate training set.

To fulfill the objectives of this study and address obscuration, particular emphasis is placed on identifying flare rise episodes rather than start-to-end profiles. Restricting the analysis to the rise episode offers multiple advantages. Most existing automated FDAs fail to detect overlapping flares as they embed the slow-driving assumption, that is, one flare must terminate before the next begins \citep{Aschwanden_2012, lu2024automatic, valluvan2024solar}. During periods of high solar activity, however, multiple flares frequently occur in close succession, and overlapping events are common and, importantly, discernible by eye. Focusing on the rise episodes partially relaxes the non-overlap constraint: while overlap is not permitted during the rise of an ongoing flare, subsequent events initiating during the decay of a preceding flare are permitted. Moreover, rise episodes are typically the shortest, so overlap is less likely, and they are the most distinct flare phases, making them straightforward to identify. In contrast, the decay phase often exhibits changes in logarithmic slope driven by different energy loss processes, which makes its end difficult to distinguish from the background.

\begin{figure*}
\centering
\includegraphics[width=\textwidth]{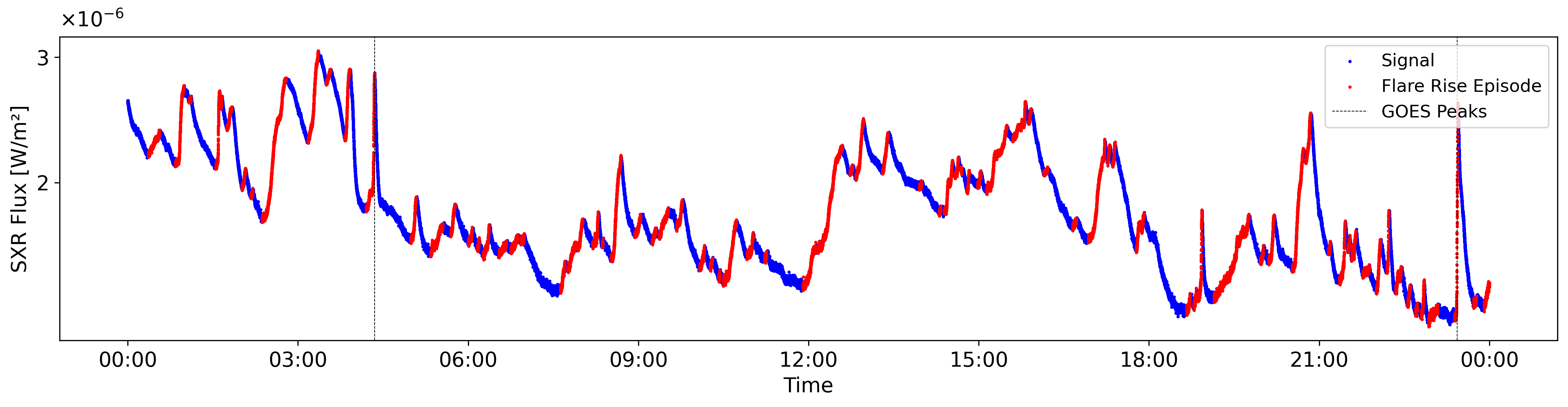}
\includegraphics[width=\textwidth]{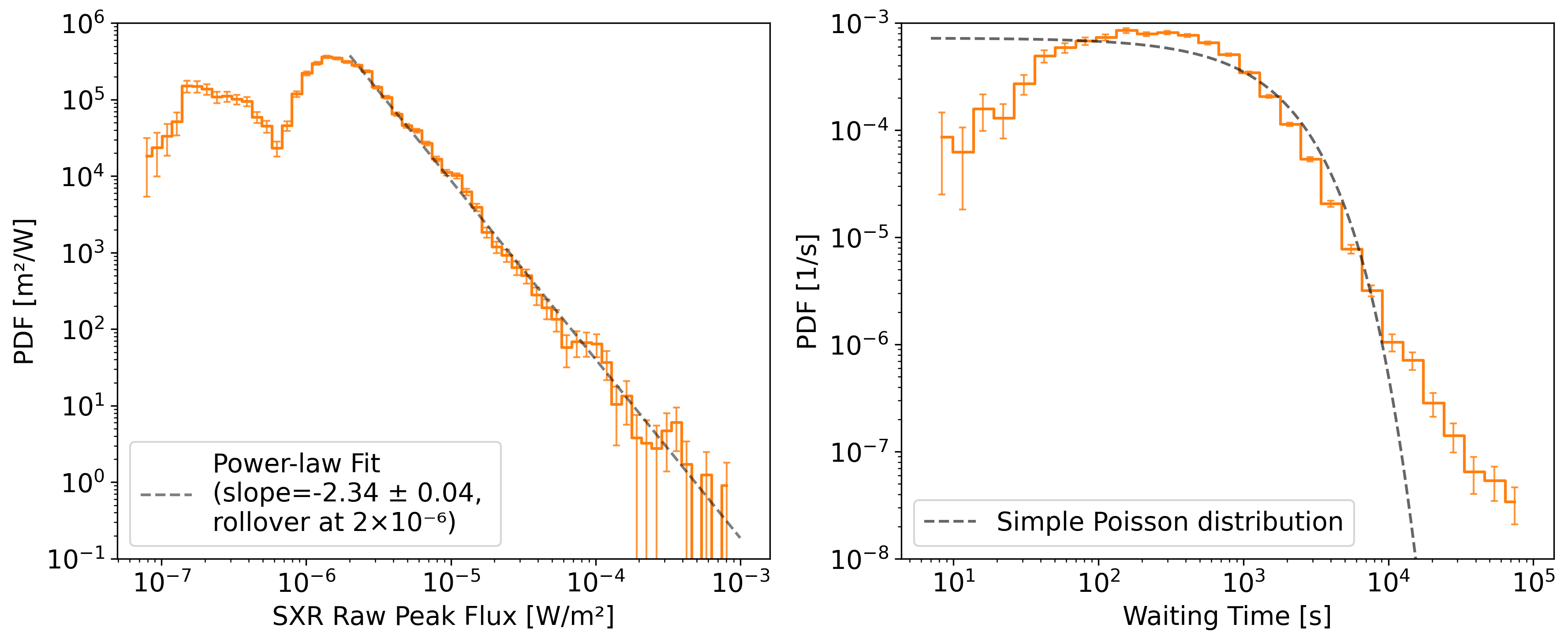}
\caption{Top panel: GOES SXR data on 13 January 2024 with 83 manually selected rise episodes marked in red, and the remaining signal shown in blue. Vertical dashed lines indicate the peak times of the two events listed in the GOES catalog. Bottom panels: PDFs of the raw peak fluxes (left panel) and waiting times (right panel) for events in the reference catalog. The reference catalog includes 145 randomly selected days between 2018 and 2025. Waiting times exceeding one day are excluded to avoid artifacts arising from data gaps.}
\label{fig_ref_cat}
\end{figure*}

The above manual procedure yields a reference catalog of 7,700 solar flares, providing the foundation for supervised training and evaluation of the CNN framework. Figure \ref{fig_ref_cat} (top panel) presents an example of GOES SXR data for 13 January 2024, with manually identified rise episodes highlighted in red. On this day, the authors manually selected 83 events, whereas the GOES catalog registered only 2. The bottom panel shows the probability distribution functions (PDFs) of the raw peak fluxes and waiting times of all events in the reference catalog. The peak fluxes follow a power-law distribution for fluxes greater than $2 \times 10^{-6}$ Wm$^{-2}$. Empirically, power laws only hold over a limited range of sizes, due to background and instrumental effects, and fitting a power law above a lower-end ``rollover'' is often more applicable \citep{aschwanden2015thresholded}. Accordingly, all power-law fits in this study are restricted to a range where the power law model applies. The waiting-time distribution exhibits a more complex behavior and departs from a simple Poisson form $P(\Delta t) = \lambda e^{- \lambda \Delta t}$, where $\lambda$ denotes the mean event rate (see Section \ref{sec:stat3}).

\subsection{Training and Validation} \label{sec:transet}
The reference catalog is used to construct the training and validation datasets. To prevent leakage, that is, the unintended use of data reserved for model evaluation within the training process \citep{de2018advances}, the catalog is partitioned by day in chronological order, with the first 80\% of the 145 days allocated to training and the remaining reserved for validation. Within each day, the GOES SXR flux signal is segmented into overlapping windows of fixed length (600 samples), termed sets. Consecutive windows are shifted by 120 samples, referred to as the stride, which corresponds to 20\% of the window size. 

Each window is normalized using \textit{z}-score standardization (i.e., the mean is subtracted and the result is divided by the standard deviation, thereby producing inputs with zero mean and unit variance). This improves numerical stability and affects how event magnitudes contribute to the learning process. Without standardization, the loss function and gradient updates would be dominated by large flares, i.e., biasing the algorithm toward optimizing predictions for the largest events while under-weighting smaller ones. Standardization ensures comparable sensitivity across the full range of flare magnitudes and enables the model to capture relative variations rather than absolute variations in keeping with the expected scale-invariant properties of solar flares. 

The corresponding label sequences are derived directly from the reference catalog: rise episodes are labeled as class 1, whereas all other intervals are assigned to class 0. This procedure yields 65,955 training and 15,727 validation sets, spanning a wide range of flare morphologies and activity levels. 

Table~\ref{tab:cnn_metrics} summarizes the CNN performance during validation and includes the definitions of the evaluation metrics. The model achieves an overall accuracy of 0.896, indicating robust performance across both classes. For rise episodes, the network attains a high recall of 0.953, showing strong sensitivity to flare rise episodes, though the corresponding precision (0.715) suggests some false detections. In contrast, the background class shows very high precision (0.983) and moderate recall (0.877), indicating that the model rarely misclassifies true rise episodes as background, yet some background intervals are falsely labeled as rises. Despite this asymmetry, the macro and weighted averages remain high (F$_1$-score $\approx$ 0.9), indicating that the model maintains balanced performance between flare and background intervals. 

High recall and precision values on validation data might indicate model overfitting, that is, the model learns noise rather than general patterns \citep{ghojogh2019theory}. While the background class attains high precision, its dominance in the dataset argues against overfitting \citep{johnson2019survey}. The imbalance between precision and recall, instead, demonstrates that the model preserves meaningful performance \citep{salehi2017tversky}.

\subsection{Application to the Full Dataset} \label{sec:testset}

The trained model is applied to the full GOES SXR dataset spanning 1 January 2018 to 22 August 2025. This period of study corresponds to 2,757 calendar days, with eleven days excluded due to missing satellite observations. For each day, the primary GOES satellite is used as the baseline. Whenever intervals are affected by data gaps or flagged as low quality, the corresponding segments from the secondary satellite are substituted, provided valid measurements are available. Over the study period, 417 days rely on data from only a single operational satellite. After substitution, any timestamps still carrying low-quality flags are discarded. As a result, most daily series contain fewer than the nominal 86,400 one-second samples. Moreover, 15 days with fewer than twelve hours of valid data are removed entirely. Test sets are then constructed from the cleaned daily series and passed to the CNN.

The model is designed to identify the flare rise episode rather than the start-to-end profile. In other words, the aim is to prevent overlaps during the rise episode of an ongoing flare, while allowing subsequent events that are initiated during the decay episode of a preceding flare. In practice, however, the CNN occasionally detects overlaps during rise episodes. Most of these correspond to spurious transitions introduced by classification error rather than genuine flares. Accordingly, overlapping detections during rise episodes are merged into a single event during post-processing, whereas detections occurring during the decay of a flare are recorded as separate events, allowing overlaps that start (or start and finish) during the decay. 

\begin{table}
\centering
\caption{Validation performance of the CNN. Precision, recall, F$_1$-score, and support are reported for both classes (rise episodes and background), along with overall accuracy, macro average, and weighted average. 
Metrics are defined as follows: 
precision $= \text{TP} / (\text{TP}+\text{FP})$, 
recall $= \text{TP} / (\text{TP}+\text{FN})$, 
F$_1 = 2 \times (\text{precision} \times \text{recall}) / (\text{precision} + \text{recall})$, 
support $= \text{TP}+\text{FN}$, 
accuracy $= (\text{TP}+\text{TN})/(\text{TP}+\text{TN}+\text{FP}+\text{FN})$, 
where TP, FP, TN, and FN denote true positive, false positive, true negative, and false negative, respectively. Macro and weighted averages are computed across the two classes. In the weighted average, each class is weighted by its support (i.e., the number of samples in that class).}
\label{tab:cnn_metrics}
\begin{tabular}{lcccc}
\hline
\textbf{Class} & \textbf{Precision} & \textbf{Recall} & \textbf{F$_1$-score} & \textbf{Support} \\
\hline
Background     & 0.983 & 0.877 & 0.927 & $7.13 \times 10^{6}$ \\
Rise episode          & 0.715 & 0.953 & 0.817 & $2.30 \times 10^{6}$ \\
\hline
Accuracy       & 0.896 & 0.896 & 0.896 & $2.30 \times 10^{-1}$ \\
Macro avg.      & 0.849 & 0.915 & 0.872 & $9.44 \times 10^{6}$ \\
Weighted avg.   & 0.918 & 0.896 & 0.900 & $9.44 \times 10^{6}$ \\
\hline
\end{tabular}
\end{table}

\subsection{A Bayesian Approach to Event Confidence}\label{sec:bayesian}
Table~\ref{tab:cnn_metrics} presents the classification performance of the CNN on the validation dataset; however, it does not assign a confidence level to individual candidates. It is important to quantify the likelihood that each CNN-detection represents a genuine solar flare, rather than noise. In principle, cross-validation against existing flare catalogs yields some information, but every catalog is incomplete. To overcome this, we introduce a Bayesian framework that enables probabilistic assessment of the reliability of each CNN detection.

\begin{figure*}
\centering
\includegraphics[width=\textwidth]{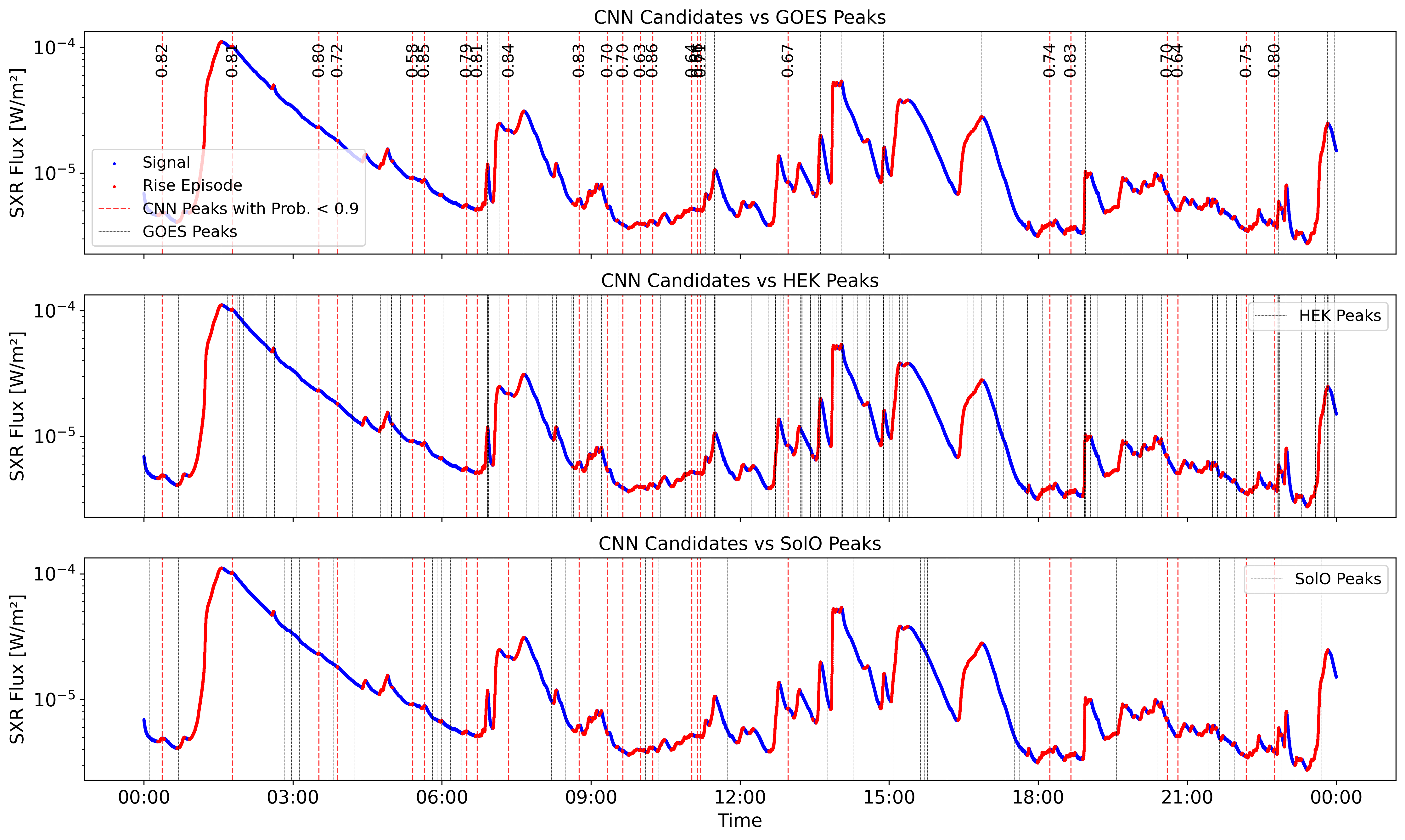}
\caption{An example of applying the CNN to GOES SXR data on 23 March 2024. Detected flare candidates (rise episodes) are marked in red, while the remaining signal is shown in blue. Peak times of events registered in the GOES, HEK, and SolO catalogs are indicated by vertical dashed gray lines in the top, middle, and bottom panels, respectively. Vertical dashed red lines mark CNN candidates below the confidence level of 0.9, and their corresponding probabilities are annotated in the top panel.}
\label{Fig1}
\end{figure*}

The classification task is posed in terms of two competing hypotheses: the null hypothesis $H_0$, corresponding to a false positive (FP), and the alternative hypothesis $H_1$, corresponding to a true positive (TP). Empirically, detections with lower energy or shorter rise times are more likely to be noise artifacts. We employ three diagnostic features: the amplitude $A$, defined as the background-subtracted peak flux; the rise time $\tau$; and a binary coincidence flag $C$, which indicates whether the candidate coincides with at least one of three existing catalogs within a tolerance window $\Delta t = 180$s. Here, the term background-subtracted denotes the peak flux minus the flux measured at the start of the rise episode, and cross-referencing for coincidence is performed against the GOES, Heliophysics Event Knowledgebase (HEK), and Solar Orbiter (SolO) catalogs (see Appendix \ref{app3}).

The prior probabilities of the hypotheses are estimated from the empirical frequencies of TPs and FPs according to the reference catalog:
\begin{align}
P(H_0) &= \frac{N_{\text{FP}}}{N_{\text{TP}} + N_{\text{FP}}}, \label{eq:ph0} \\
P(H_1) &= \frac{N_{\text{TP}}}{N_{\text{TP}} + N_{\text{FP}}}. \label{eq:ph1}
\end{align}
The likelihood function of the observed variables $(\tau, A, C)$ conditioned on hypothesis $H_{i}$ is then defined as
\begin{equation}
P(\tau, A, C \mid H_{i}) 
= P(A \mid H_{i}) \, P(\tau \mid A, H_{i}) \, P(C \mid H_{i}),
\label{eq:bayes_features}
\end{equation}
where $P(A \mid H_i)$ denotes the marginal distribution of amplitude conditioned on $H_i$, obtained empirically from the histogram of amplitudes corresponding to TPs or FPs, while disregarding rise-time information. The conditional distribution $P(\tau \mid A, H_i)$, which accounts for the dependence between amplitude and rise time, is estimated by projecting the joint scatter of amplitude and rise time onto the amplitude axis for each hypothesis. Specifically, for a given amplitude $A$ with the arbitrary choice of $dA = 0.1 A$, all rise time values within the interval $[A - dA, A + dA]$ are selected to construct a histogram that is then normalized to yield $P(\tau \mid A, H_i)$. This formulation captures the observed tendency for stronger events to exhibit longer rise times.

The coincidence flag is modeled as a deterministic binary observation. A value of $C=1$ indicates the existence of at least one catalog event within $\pm \Delta t$ of the candidate, whereas $C=0$ signifies the absence of such a coincidence. This observation strongly modulates the posterior inference. For $C=1$, the posterior probabilities collapse to certainty in favor of the TP hypothesis, that is $ P(H_1 \mid A, \tau, C=1) = 1$
and $P(H_0 \mid A, \tau, C=1) = 0$.
For $C=0$, the posterior probability relies exclusively on the joint distribution of amplitude and rise time. The probability of a TP is given by Bayes’ theorem as
\begin{equation}
P(H_1 \mid A, \tau, C=0) =
\frac{P(A, \tau \mid H_1) \, P(H_1)}
     {P(A, \tau \mid H_1) \, P(H_1) + P(A, \tau \mid H_0) \, P(H_0)},
\label{eq:posterior_no_coincidence}
\end{equation}
with the complementary probability
\begin{equation}
P(H_0 \mid A, \tau, C=0) = 1 - P(H_1 \mid A, \tau, C=0),
\label{eq:posterior_h0}
\end{equation}
quantifying the confidence that the event constitutes a spurious detection. 

Figure~\ref{Fig1} shows an example of applying the CNN to GOES SXR data on 23 March 2024. The probability of each identified event is quoted in the top panel. The figure illustrates that CNN identifies numerous flare candidates, including events absent from the GOES, HEK, and SolO catalogs. Further details and discussion of this figure are provided in Section~\ref{res:conf}.

\section{A New Catalog} \label{sec:res}
The catalog is constructed for 1 January 2018 to 22 August 2025. Over this interval, the CNN identifies 111,580 flaring candidates, compared with 14,612 events recorded in the GOES catalog for the same period. In this section, we quantify the confidence of the detected candidates (Section \ref{res:conf}), investigate the impact of day boundaries on event detections (Section \ref{sec:mdf}), evaluate the statistical properties of events (Sections \ref{sec:stat1} and \ref{sec:stat2}), and calculate their size and waiting time PDFs (Section \ref{sec:stat3}). Throughout, we compare the measured statistics with those from the GOES catalog, which shares the same temporal coverage and SXR observations, and serves as the standard reference in solar flare studies. In Section \ref{res:cats_comp}, We put the CNN catalog in context with the catalogs summarized in Appendix~\ref{app3}. We deliberately follow a conservative approach and compare the catalogs in terms of total event counts in the first instance. More elaborate comparisons, e.g. based on physical properties or one-to-one event matching, are postponed to future work, as it is unclear at the time of writing how to correct systematically for differences in detection criteria, observing bands, sensitivity thresholds, and temporal coverage between different catalogs.

\subsection{Detection Confidence} \label{res:conf}
Figure~\ref{Fig1} presents the results of applying the CNN to the GOES SXR flux on 23 March 2024. The detected rise episodes are indicated in red. Vertical gray lines mark the peak times reported in three independent flare catalogs: GOES (top panel), HEK (middle panel), and SolO (bottom panel). On this date, the number of events recorded in the GOES, HEK, SolO, and CNN catalogs are 19, 230, 64, and 94, respectively. For each CNN candidate, the probability of being a TP rather than a FP is quantified using the Bayesian framework introduced in Section \ref{sec:bayesian}. In all panels, CNN candidates with posterior probability (confidence level) below 0.9 are highlighted by red dashed vertical lines, and their corresponding probabilities are annotated in the top panel. Figure~\ref{Fig1} illustrates how the classification scheme operates on a single day of data: among the 94 CNN detections, 70 are classified as TPs and 24 as FPs when a confidence level of 0.9 is applied. The FPs typically correspond to weak or short-lived fluctuations. Similar results are obtained when the scheme is applied to the entire period of study.

Table \ref{tab_2} summarizes the number of CNN detections classified as TP at different confidence levels for the whole catalog. Overall, at the high confidence level of 0.9, 62,896 out of 111,580 CNN candidates are classified as TPs, that is, over four times the number of events listed in the GOES catalog. Using a more permissive level of 0.2 results in nearly all CNN candidates (111,486) being classified as TPs, making the CNN catalog more than seven times larger than the GOES catalog. In the rest of this paper, we analyze events with $P \geq 0.2$. Upon checking a higher threshold, $P \geq 0.5$, we find no significant changes in the resulting statistics; for example, the peak flux histograms are barely distinguishable by eye from those in Figure \ref{Fig3}.

\begin{table} 
\centering
\caption{Number of CNN candidate events retained at different confidence levels. Percentages are expressed relative to the total number of CNN candidates (111,580).}
\begin{tabular}{c c c c}
\hline
\textbf{Confidence Level} & \textbf{No. TPs} & \textbf{No. FPs} & \textbf{Percentage of TPs}\\
\hline
$P > 0.2$ & 111,486 & 94 & $\approx$100\% \\
$P > 0.5$ & 91,502  & 20,078 & 82\%  \\
$P > 0.9$ & 62,896  & 48,684 & 56\%  \\
\hline
\label{tab_2}
\end{tabular}
\end{table}

\subsection{Detections Across Day Boundaries} \label{sec:mdf}
The algorithm processes each day independently, which can affect the identification of flare rise episodes that span two consecutive days. In practice, there are no records of events ending exactly at 23:59:59 and new events starting exactly at 00:00:00 on the following day in the CNN catalog. For this reason, we examine two narrow time windows bracketing the day boundary: one spanning 23:59:00--23:59:59 on a given day and another spanning 00:00:00--00:01:00 on the following day. We then search for event peak times falling within the former window and event start times falling within the latter.

We find overall that detections across day boundaries are rare and do not affect the event statistics significantly. Specifically, when the dominant portion of the rise occurs on the first day and extends weakly into the following day, the rise onset is correctly identified, but the peak time is assigned to the end of the first day rather than to its true occurrence on the following day. We identify 344 events of this type in the CNN catalog. On the other hand, when a flare begins near the end of the first day and intensifies on the following day, the onset is not detected until the second day, delaying the inferred start time even though the peak time is correctly identified. This behavior is observed for 123 events in the CNN catalog. Finally, on rare occasions, the algorithm identifies rise phases on both days. We find only 6 such cases in the CNN catalog. Visual inspection indicates that they correspond to genuinely separate flares rather than duplicate detections of a single event.

\subsection{Event Counts Across Flare Classes} \label{sec:stat1}
\begin{figure}
\centering
\includegraphics[width=\columnwidth]{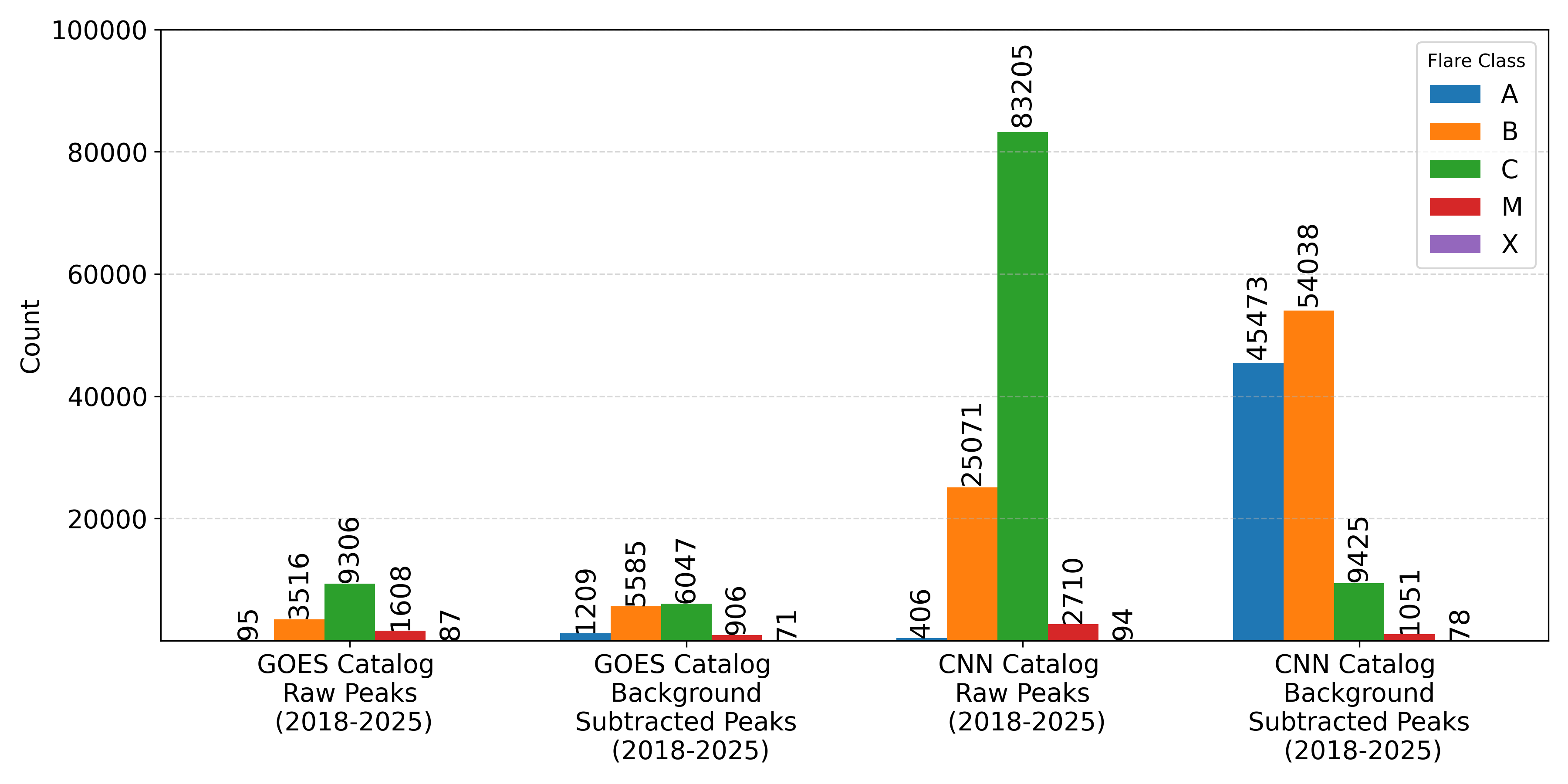}
\caption{Event count per flare class for GOES (left two histograms) and CNN (right two histograms) catalogs. Flare classes are color-coded according to the legend. For each catalog, separate histograms are graphed for raw and background-subtracted peak fluxes.}
\label{Fig2}
\end{figure}

It is important to compare how effectively the new CNN catalog and the GOES catalog detect flares of various classes. Figure \ref{Fig2} compares the distribution of events classified according to the GOES system, viz. classes A, B, C, M, and X. For both catalogs, we plot separate histograms for the event classes derived from raw and background-subtracted peaks. The background subtraction is applied after the CNN has identified events. Applying it before the CNN detection would require a detailed characterization of the background flux, which lies beyond the scope of this study.

Figure~\ref{Fig2} shows that the CNN catalog captures a substantially larger population of small events compared with the GOES catalog. This arises from the CNN’s design, which targets the flare rise episodes rather than full start-to-end durations and resolves overlapping events. For raw peak fluxes, the number of A-, B-, and C-class events increases significantly from 95, 3,516, and 9,306 in the GOES catalog to 406, 25,071, and 83,205 in the CNN catalog. The enhancement extends to stronger flares as well, with M-class events increasing from 1,608 to 2,710 and X-class events from 87 to 94. 

When the background contribution is removed, the distribution of events shifts markedly toward smaller flare classes. For example in the CNN catalog, the number of A- and B-class events increases from 406 and 25,071 to 45,473 and 54,038, respectively, while the number of C-class events decreases from 83,205 to 9,425. The counts of larger flares also decline from 2,710 to 1,051 for M-class and from 94 to 78 for X-class events. This redistribution highlights the importance of background subtraction for event classification. Overall, the results imply that the CNN adds resolution at the low peak-flux end of the distribution without artificially inflating the counts of large flares. 

\subsection{Peak Flux Over a Solar Cycle} \label{sec:stat2}
\begin{figure}
\centering
\includegraphics[width=\columnwidth]{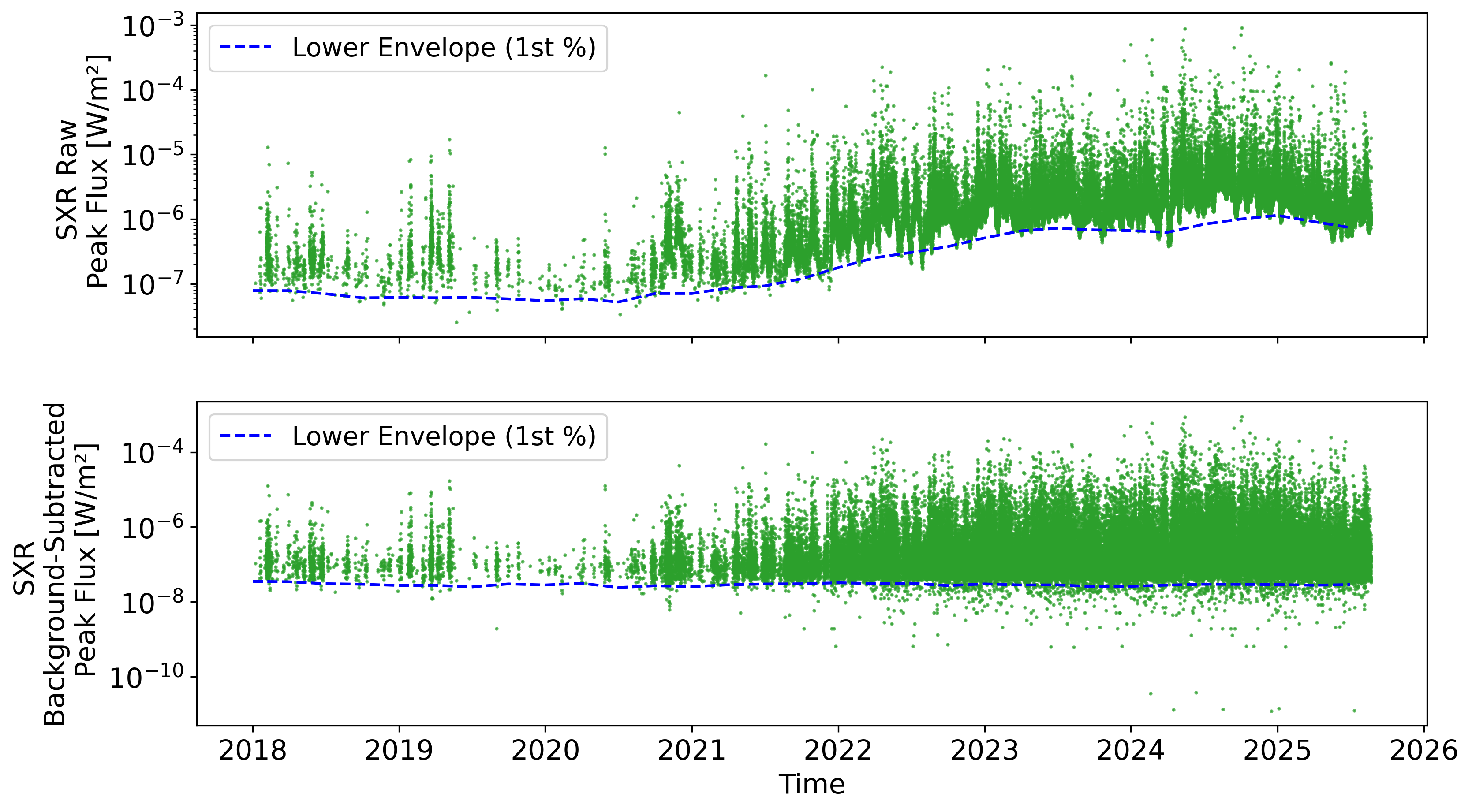}
\caption{Raw (top panel) and background-subtracted (bottom panel) peak fluxes versus time for events in the CNN catalog. The blue dashed lines mark the lower envelope of event sizes, corresponding to the 1st \% quantile of peak fluxes within each three-month period.}
\label{fig_flux}
\end{figure}

The periodic modulation of flare peak fluxes over the solar cycle is well known \citep{aschwanden1994irradiance, kopp2025solar}. It is natural to ask whether the phenomenon manifests itself in the CNN catalog. Figure \ref{fig_flux} shows the CNN-identified peak fluxes versus time over the full study interval, with raw peaks in the top panel and background-subtracted peaks in the bottom panel. The raw peak fluxes display the expected modulation with the solar cycle and, in particular, show a time-dependent minimum event size (the lower envelope of the scattering of events). The lower envelope is defined as the value below which the lowest 1\% of the peak fluxes fall, averaged within each three-month period. These values are then smoothed over three consecutive periods. The lower envelope exhibits a gradual increase, from solar minimum to solar maximum, in flux level by 1.3 orders of magnitude, as determined from the logarithmic ratio of the maximum to minimum values of the envelope. Similarly, there is an increase in the number of events per month by about 3.2 orders of magnitude, derived from the ratio of the maximum to minimum monthly flare counts over the solar minimum in 2018 to the solar maximum in 2025. The trend reflects the progressive elevation of the SXR background, as ARs become more numerous and magnetically complex from 2018 to 2025. The bottom panel shows that background subtraction effectively removes the time dependence of the minimum event size.

\subsection{Size and Waiting-Time Distributions}\label{sec:stat3}
The PDF of solar flare peak fluxes is often characterized as a power law over several orders of magnitude \citep{Mike2008, aschwbook2011, Aschwanden2014, Farhang_2022}, consistent with scale-free and self-organized critical processes \citep{aschwanden201625, farhang2018, kaki2022evidence}. In contrast, the statistical properties of flare waiting times remain uncertain. Several studies reported that waiting times broadly exhibit homogeneous or inhomogeneous Poisson-like statistics, implying a memoryless process, but the exact functional form of the waiting-time distribution depends on the temporal and spatial scope of analysis \citep{wheatland1998waiting, Wheatland2000, Litvinenko2002, farhang2019, aschwanden2021poissonian}. However, some authors report departures from Poisson behavior, suggesting correlations or memory between events \citep{lepreti2001solar, li2018waiting, verbeeck2019solar, lei2020solar, Hudson2020, Carlin2023}. Here, we revisit the above issues and assess whether they are influenced by the detection limitations of existing FDAs compared to the CNN framework.

\begin{figure*}
\centering
\includegraphics[width=0.48\textwidth]{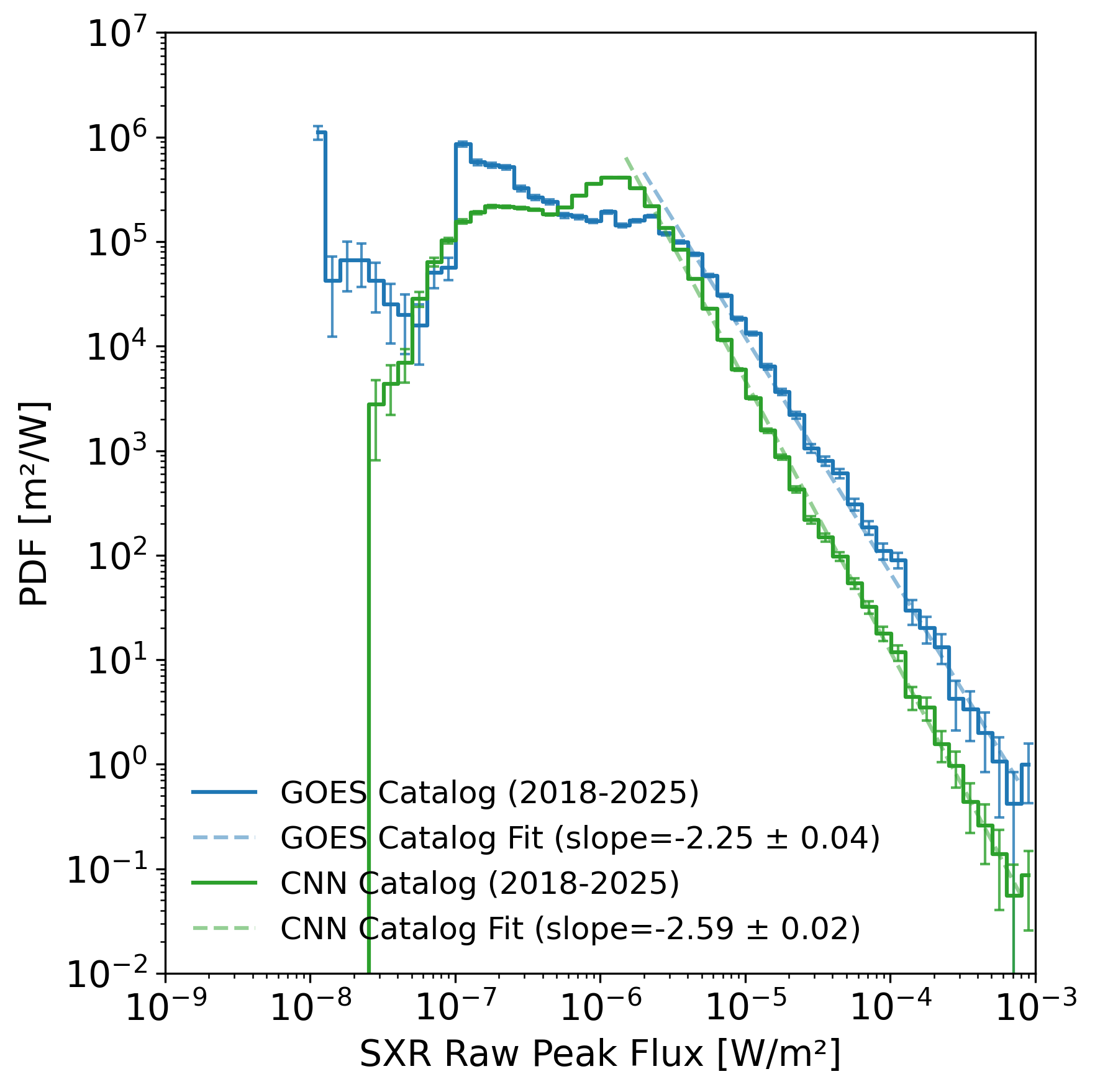}
\hfill
\includegraphics[width=0.48\textwidth]{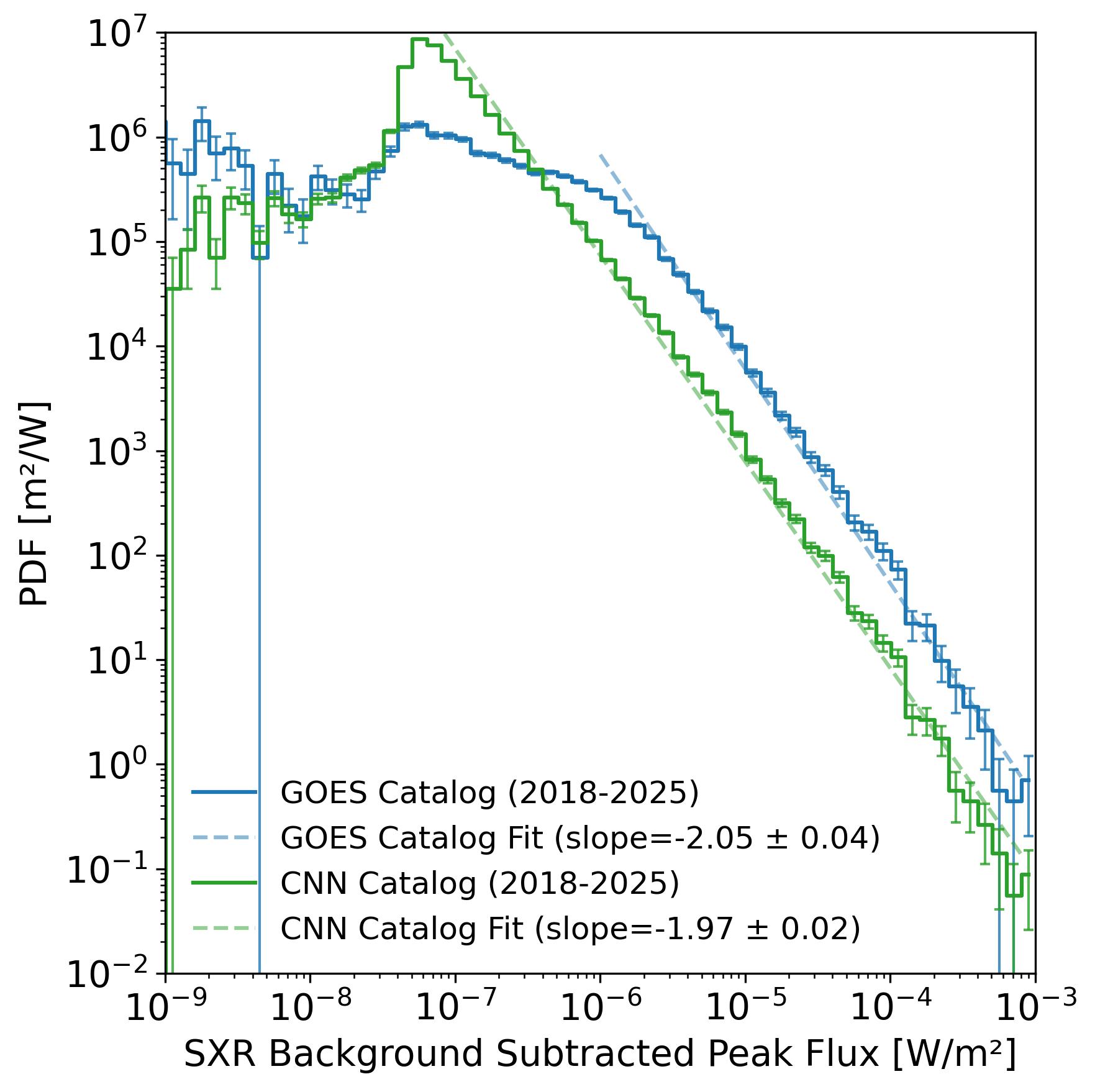}
\caption{Flare peak flux distribution: PDF of raw (left panel) and background-subtracted (right panel) SXR peak flux for the GOES (blue histograms) and CNN (green histograms) Catalogs. The dashed curves indicate the fitted power laws over the applicable ranges. For ease of comparison, the left and right scales are the same.}
\label{Fig3}
\end{figure*}

Figure~\ref{Fig3} displays the PDFs of the raw (left panel) and background-subtracted (right panel) peak fluxes for the GOES (blue histograms) and CNN (green histograms) catalogs. The figure demonstrates that the peak distribution retains a power-law form in both GOES and CNN catalogs above a lower-end rollover. The CNN distribution exhibits a slightly steeper power-law index in the raw peak fluxes ($-2.59 \pm 0.02$) compared to that of the GOES catalog ($-2.25 \pm 0.04$), reflecting the inclusion of a larger population of small events. Once the background contribution is removed, the two distributions exhibit more similar power-law indices of $-2.05 \pm 0.04$ and $-1.97 \pm 0.02$ for the GOES and CNN catalogs, respectively. A slight concavity is also apparent in both distributions, when they are plotted on a log-log scale, which may arise from catalog incompleteness. The convergence indicates that the CNN detections are not the result of overfitting the background noise, because if spurious detections are common, the power-law index of the background-subtracted CNN catalog would deviate from that of the background-subtracted GOES catalog. Furthermore, the new catalog with background subtraction extends the power-law range by about an order of magnitude (spanning $10^{-7}$ to $10^{-3}$ Wm$^{-2}$) compared with the GOES catalog.

\begin{figure}
\centering
\includegraphics[width=\columnwidth]{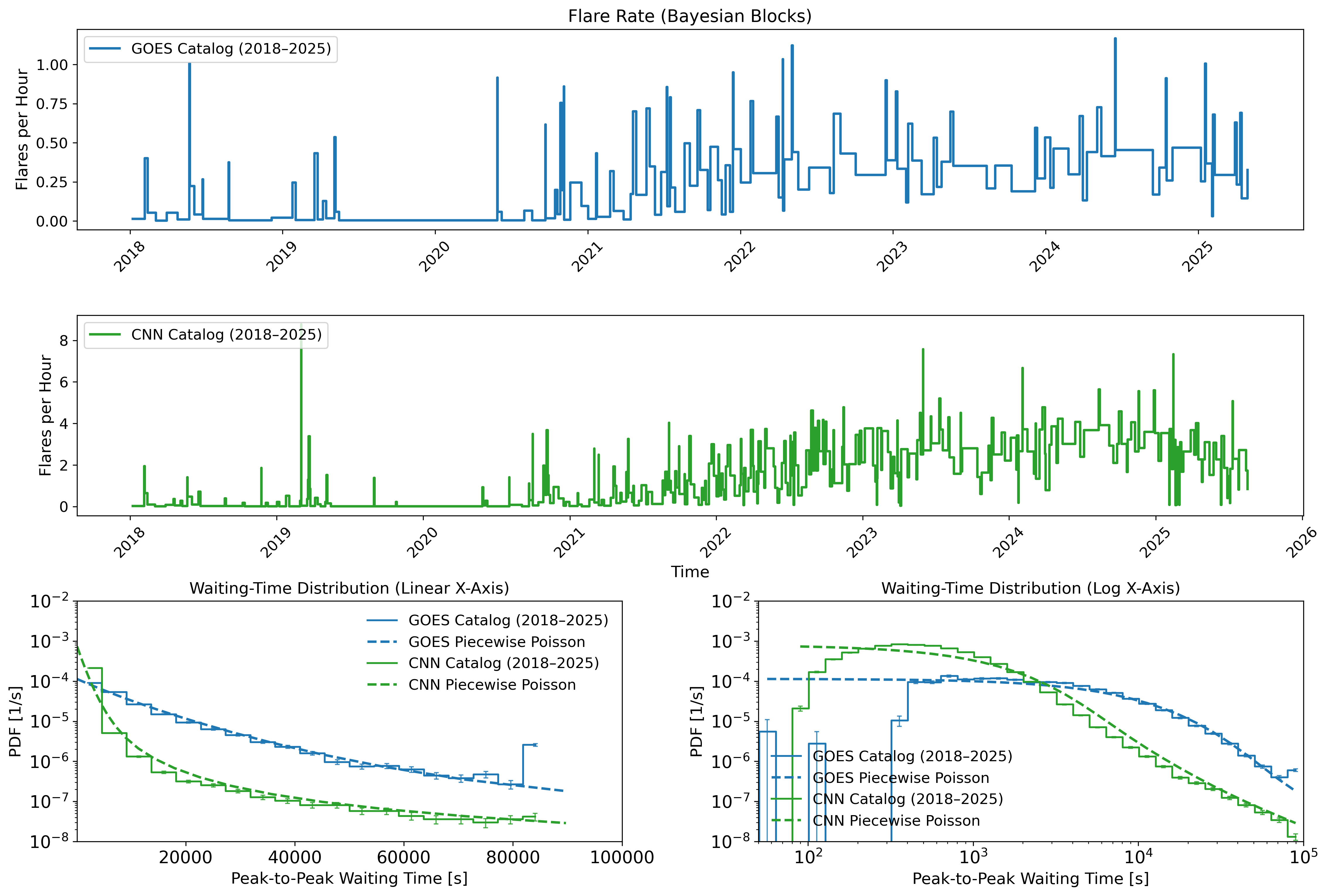}
\caption{Flare waiting time statistics. Top and middle panels: flare rates derived from applying the Bayesian block algorithm to the waiting-time distributions of the GOES (blue) and CNN (green) catalogs. Bottom panels: waiting-time distributions for both catalogs, along with the piecewise Poisson models of Equation (\ref{poisson}), plotted on linear-log (left) and log-log (right) scales.}
\label{fig_rate}
\end{figure}

Next, we examine the peak-to-peak waiting times ($\Delta t$) of events for both the GOES and CNN catalogs. We consider a piecewise constant Poisson model \citep{mike2001} for the waiting-time distributions,
\begin{equation}
P(\Delta t) = \sum_i \phi_i \lambda_i e^{-\lambda_i \Delta t},
\label{poisson}
\end{equation}
with $\phi_i = \dfrac{\lambda_i t_i}{\sum_{i} \lambda_i t_i}$, where $\lambda_i$ denotes the flare rate over the interval $t_i$. The intervals $t_i$ are determined using the Bayesian Blocks algorithm, which partitions the time series into segments and optimizes the number and boundaries of segments, such that the event rate remains approximately constant within each interval \citep{scargle2013studies}. The obtained flare rates are plotted as a function of time in Figure \ref{fig_rate} for the GOES (top panel) and CNN (middle panel) catalogs. The CNN catalog reveals periods of higher rate, reaching up to about 8 flares per hour compared with 1 flare per hour for GOES. The weighted mean rate, $\langle \lambda \rangle = \frac{\sum t_i \lambda_i}{\sum t_i}$, is $1.35 \pm 1.43$ flares per hour for the CNN catalog and $0.21 \pm 0.21$ flares per hour for GOES, where the quoted uncertainties represent the standard deviations. The averages and standard deviations account for the varying lengths of the Bayesian blocks. The standard deviations are comparable to or exceed the mean rates, indicating strong temporal variability in flare occurrence. The elevated flare rate observed in the CNN reflects its enhanced sensitivity to detecting small events.

The waiting-time distributions of both the GOES (shown in blue) and CNN (shown in green) catalogs, along with the corresponding piecewise constant Poisson models, are displayed in the bottom panel of Figure \ref{fig_rate}. The distributions are shown on linear-log axes (left panel), where an exponential waiting time corresponding to a constant-rate Poisson process would appear as a straight line, as well as log-log axes (right panel). The relative differences between the observed and modeled PDFs are characterized by the fractional deviation, defined as \(\delta = [P_{\mathrm{obs}} - P_{\mathrm{model}}] / P_{\mathrm{model}}\) with \(\langle \delta \rangle\) denoting the mean fractional deviation, and \(\langle |\delta| \rangle\) the mean absolute fractional deviation, both averaged over all bins. Excluding the last anomalous bin ($\Delta t > 80,000$)s, the GOES distribution yields \(\langle \delta \rangle = -0.2\%\) and \(\langle |\delta| \rangle = 15.6\%\), indicating consistency with the piecewise Poisson model. The CNN distribution exhibits a larger departure, with \(\langle \delta \rangle = -17.1\%\) and \(\langle |\delta| \rangle = 26.4\%\). Both PDFs remain broadly consistent with a piecewise Poisson process, although the CNN diverges slightly more. We note that the above approach quantifies the discrepancy between the distributions and the piecewise Poisson model under the assumption that the model and its parameters are exact, and only accounts for the variability in $P_\text{obs}$, which reflects sampling noise (e.g., observational uncertainty, finite statistics, or binning effects). To incorporate model uncertainties, a resampling technique such as bootstrapping could be employed \citep{crowley2024superflare}.

\begin{table}[ht]
\centering
% \caption{Number of flares registered in different catalogs.}
\caption{Number of flares registered in different catalogs.}
%: the CNN catalog, the GOES catalog, the Reuven Ramaty High Energy Solar Spectroscopic Imager (RHESSI) catalog, the Hinode, the Aschwanden \& Freeland catalog, the Temperature and Emission Measure-Based Background Subtraction (TEBBS) catalog, the Solar Photometer in X-rays (SphinX) catalog, the Solar Orbiter (SolO) catalog, the Fermi Large Area Telescope (Fermi-LAT) catalog, the Convolutional Physics-based Hierarchical Learning for AR Events (C-PhLARE) catalog, the AGILE catalog, and the Heliophysics Event Knowledgebase (HEK) catalog.}

\begin{tabular}{lcccc}
\hline
Catalog & Analysis Interval & No. Flares \\
\hline
CNN  &  2018--2025  & 111,580 \\ 
GOES (\ref{A1})  &  2018--2025  & 14,612 \\
  % &  (1975--present) &  \\
RHESSI (\ref{A2}) & 2002--2018 & 104,036 \\
Hinode (\ref{A3}) & 2006--2025 & 30,219 \\
Aschwanden \& Freeland (\ref{A4})  & 1975--2011 & 338,661 \\
TEBBS (\ref{A5}) & 1980--2007 & 50,000 \\
SphinX (\ref{A6}) & 2009 & 1,604 \\
SolO (\ref{A8}) & 2020--2025  & 79,526 \\
Fermi-LAT (\ref{A9})  & 2010--2018  & 45 \\
C-PhLARE (\ref{A10})  & 2011--2018 & 18,833 \\
AGILE  (\ref{A11}) & 2007--2022 & 5003 \\
HEK (\ref{A13})  &  2018--2025  & 112,323 \\
  % &  (2012--present) &  \\
 
\hline
\end{tabular}
\label{tab:catalog_comparison}
\end{table}

\subsection{Summary Comparison of Event Counts among Catalogs}\label{res:cats_comp}
To situate the CNN catalog in a broader context, we compare it with previous catalogs. We deliberately take a conservative approach and adopt event counts as a relatively unambiguous basis for the comparison, as set out in Table \ref{tab:catalog_comparison}. Existing catalogs are highly heterogeneous with respect to detection criteria, confidence metrics, and observational conditions. As a result, even seemingly straightforward characteristics (such as minimum, maximum, or average event sizes and waiting times) are not directly comparable across catalogs, as they probe different passbands, sensitivity thresholds and temporal coverage. The event counts in Table \ref{tab:catalog_comparison} correspond exclusively to the analysis intervals listed in the table, rather than to the total interval covered by each catalog. Further details on the construction of each catalog, including the FDAs employed, instrumentation, observing wavelengths, and total observing interval, are provided in Appendix \ref{app3}.

Over the years 2018--2025, the GOES catalog lists 14,612 flares detected at 1--8\,\AA\ using the NOAA FDA, while the Heliophysics Event Knowledgebase (HEK) catalog lists 112,323 events using EUV observations. Other X-ray-based catalogs report comparable or larger event counts depending on cadence, sensitivity, and observing time, including 104,036 events in the Reuven Ramaty High Energy Solar Spectroscopic Imager (RHESSI) catalog (2002--2018), 30,219 events in the Hinode catalog (2006--2025), 50,000 events in the Temperature and Emission Measure-Based Background Subtraction (TEBBS) catalog (1980--2007), 1,604 events in the Solar Photometer in X-rays (SphinX) catalog (2009), and 79,526 events in the Solar Orbiter (SolO) catalog (2020--2025).

The Aschwanden \& Freeland catalog, constructed from background-subtracted GOES SXR data, reports 338,661 events during 1975--2011, nearly five times the number of GOES events over the same period. The Colorado Physics Laboratory Academic Research Effort Collaboration (C-PhLARE) catalog also uses GOES SXR data, applies an advanced background-subtraction procedure, and investigates the impact of different pre-flare background thresholds on the number of detected events. During 2011–2018, the catalog reports 18,833 events. 

High-energy catalogs from the First Fermi-Large Area Telescope (Fermi-LAT) and AGILE contain fewer events (45 and 5003, respectively), reflecting their sensitivity to only the most energetic flares and their more restrictive detection conditions.

\section{Obscuration} \label{sec:obscur}
Whether solar flares occur as independent, spontaneous, spatially uncorrelated events or as part of a coordinated spatially correlated stress–relaxation cycle remains an open question. One plausible explanation is that magnetic stress gradually accumulates in an AR until an instability triggers an energy release \citep{priest2002magnetic}. In this perspective, more energetic flares are expected to require longer stress accumulation, implying a positive correlation between flare sizes and backward waiting times since the preceding event \citep{hudson2020buildup, Carlin2023}. An alternative view is that the system remains close to instability at all times, and individual flares, regardless of their magnitude, produce only minor relaxation, as suggested by avalanche models \citep{aschwanden2019nonstationary}. In this view, flare sizes and waiting times are not expected to correlate, as energy release occurs spontaneously within a system which is always close to instability \citep{norman2001waiting, aschwanden2011self}. Observational studies have found little \citep{Hudson2020} or no correlation between flare sizes and waiting times \citep{Wheatland2000, Lippiello2010}. The most direct way to assess whether magnetic stress accumulation controls flare timing is through analyses of event statistics within individual ARs, where local magnetic conditions and energy input rates can be tracked explicitly. This will be the focus of a future study. In the present work, we examine the statistical behavior of flares aggregated over the entire solar disk, providing a global view of the flaring process.

\begin{figure*}
\centering
\includegraphics[width=\textwidth]{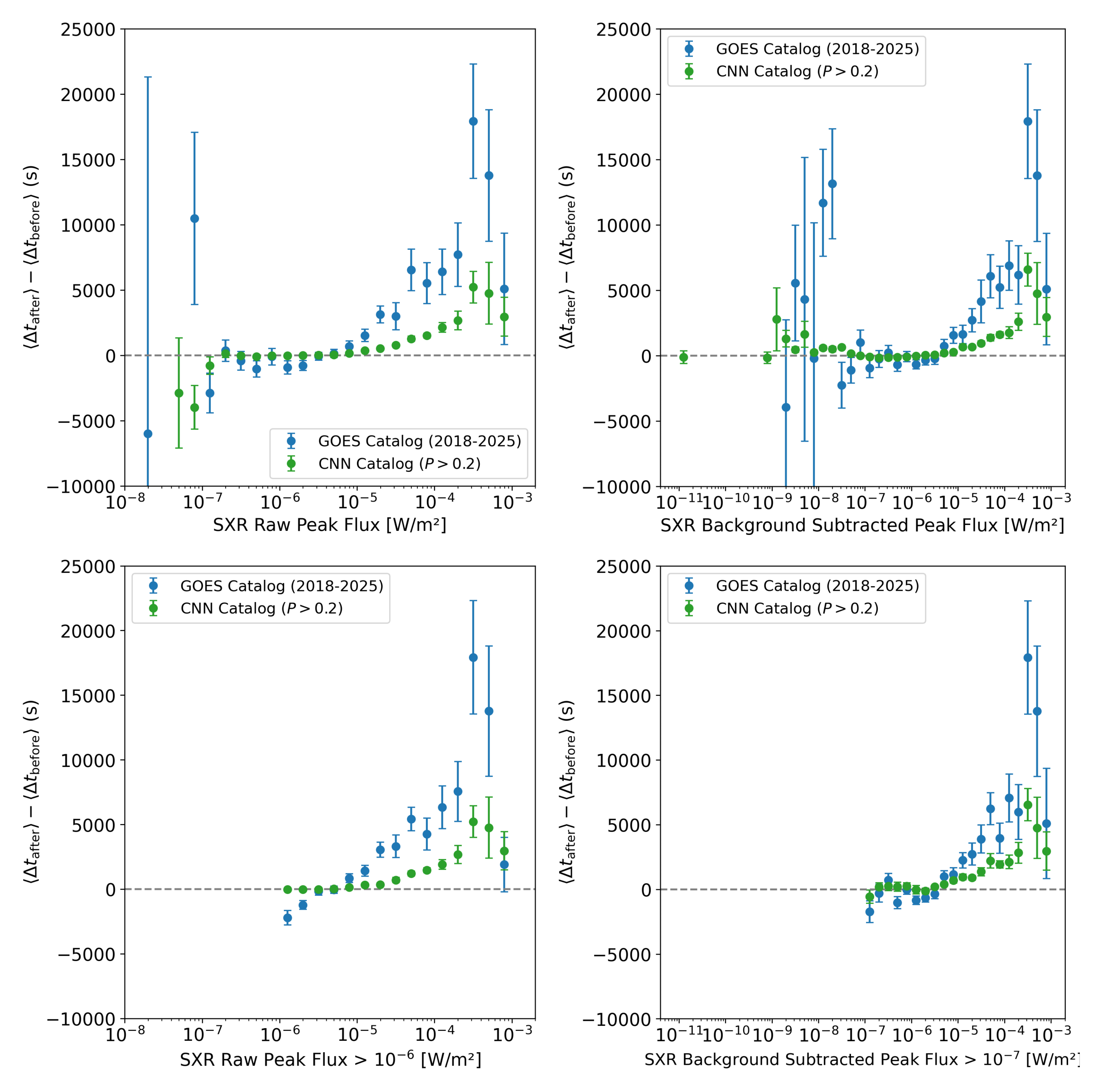}
\caption{Difference between mean waiting times after and before an event, as a function of SXR peak flux. The left and right columns show results for raw and background-subtracted peak fluxes, respectively. The top row includes all events, while the bottom row is restricted to events with peak flux above some threshold. Blue and green markers represent GOES and CNN data, respectively.}
\label{Fig5}
\end{figure*}

Figure \ref{Fig5} graphs the difference between the mean waiting time after and before an event as a function of peak flux for the GOES and CNN catalogs. Results are presented for two event selections: the full catalog (top row) and a subset limited to events with peak fluxes greater than some threshold (bottom row). Applying an intensity threshold, a common approach in scaling studies, reduces biases associated with sample incompleteness at low intensities \citep{corral2004long, baiesi2006intensity, laurson2009effect}. By examining both event selections, therefore, one can assess how the under-representation of weak events impacts waiting-time statistics. The thresholds for the raw and background-subtracted peak fluxes are set to $10^{-6}$ and $10^{-7}$ Wm$^{-2}$, respectively, above which the peak flux distributions follow a power law (see Figure~\ref{Fig3}). For each selection, waiting times are computed separately before and after each event, and mean values are evaluated within each flux bin. 

Figure \ref{Fig5} tests whether the Sun ``remembers'' its most recent flare, in the sense that a large flare is followed by a longer or shorter interval than the one preceding it. Results are shown for both raw (left panels) and background-subtracted (right panels) peak fluxes. In the top row, the GOES catalog (blue markers) shows a positive trend for larger flares, indicating a slight time asymmetry, whereby larger events are followed by longer waiting times. By contrast, the CNN catalog (green markers) exhibits a smaller asymmetry, with mean intervals before and after flares being nearly equal. This discrepancy suggests that the apparent asymmetries arise from catalog incompleteness rather than reflecting a physical suppression of flare activity following large events. Similar deviations from symmetry are observed in the bottom panels.

Figure \ref{Fig5} support the view that solar flares constitute an ongoing, random process of energy release when viewed as events from the whole Sun. The asymmetry seen in the GOES record can be interpreted as an artifact of catalog incompleteness due to obscuration, that is, the under-counting of small, temporally adjacent events during periods of elevated flux. By recovering these missed events, the apparent asymmetry is reduced, indicating that the occurrence of a flare weakly influences the timing of subsequent events.

% \begin{figure*}
% \centering
% \includegraphics[width=0.48\textwidth]{F5.png}
% \hfill
% \includegraphics[width=0.48\textwidth]{F6.png}
% % \includegraphics[width=0.48\textwidth]{R1F5.png}
% % \hfill
% % \includegraphics[width=0.48\textwidth]{R1F6.png}
% \caption{Difference between mean waiting times after and before an event versus raw (left panel) and background-subtracted (right panel) peak fluxes. Blue and green markers represent GOES and CNN data, respectively.}
% \label{Fig5}
% \end{figure*}

\section{Conclusion} \label{sec:conclusion}
A CNN–based framework is developed to generate a solar flare catalog directly from high-resolution GOES SXR data. The method complements and extends the GOES flare catalog by improving sensitivity to small and temporally adjacent events during periods of elevated flux. The CNN is trained to identify flare rise episodes rather than start-to-end profiles, as the former represent the more distinct and reliably detectable portion of flares. This design avoids uncertainties associated with determining flare end times and allows enhanced detection of overlapping flares.

The resulting catalog, covering 1 January 2018 to 22 August 2025, includes 111,580 events, which is more than seven times the number of events in the GOES catalog. A Bayesian framework is developed to quantify the probability that a CNN detection is a true positive. The approach integrates the joint probability distribution of background-subtracted peak fluxes and rise times with temporal coincidences between CNN events and the GOES, HEK, and SolO catalogs. It is found that even at a high confidence level of $P > 0.9$, the catalog retains 62,896 events.

The PDFs of the raw and background-subtracted peak fluxes of the CNN catalog are examined and compared with the GOES catalog. The PDFs are found to follow a power law above a lower-end rollover. For raw peak fluxes, the CNN catalog yields a power-law index of $-2.59 \pm 0.02$, compared with $-2.25 \pm 0.04$ for the GOES catalog. For the background-subtracted peak fluxes, the two catalogs exhibit nearly similar indices of $-1.97 \pm 0.02$ for the CNN and $-2.05 \pm 0.04$ for the GOES catalog. Moreover, the background-subtracted CNN catalog extends the detectable power-law regime by an additional order of magnitude at the lower end, spanning $10^{-7}$ to $10^{-3}$ Wm$^{-2}$.

The waiting times of the CNN events are compared with those of the GOES catalog to determine whether the arbitrary flare definition and the slow-driving assumption in the NOAA GOES FDA (also present in several other FDAs, limiting their sensitivity to detecting small or overlapping events) bias the derived statistics and the inferred physical interpretations. This exercise adds clarity to the continuing debate regarding the existence of a correlation between flare sizes and waiting times, reflecting whether solar flares are spatially uncorrelated processes governed by self-organized criticality or result from spatially correlated stress–relaxation dynamics. Applying the Bayesian block algorithm reveals that the waiting time PDFs of both the GOES and CNN catalogs are broadly consistent with a piecewise Poisson process, with an approximately memoryless character. The mean flare rate of the CNN catalog is markedly higher (1.35 $\pm$ 1.43 h$^{-1}$) than that of the GOES catalog (0.21 $\pm$ 0.21 h$^{-1}$). Further examination of pre- and post-event waiting times indicates that earlier reports of correlations between flare sizes and waiting times arise largely from obscuration, that is, under-counting weaker or overlapping flares during periods of elevated flux. When the missing events are counted by the CNN, flare occurrence is more nearly time-symmetric, consistent with a memoryless energy-release process. A complementary analysis of event statistics within individual ARs would provide an even more rigorous test of the time-symmetry and memoryless property and opens an interesting avenue for future work.

The new FDA opens a path to future studies, including mapping identified events to their source ARs by incorporating EUV observations, quantifying flare statistics within individual ARs, and examining the relationship between event size and waiting time in the broader context of flare triggering.

\section*{Acknowledgments}
The authors gratefully acknowledge the use of GOES SXR data provided by the NOAA. We thank Janet Machol for the insightful discussions on data. We also thank the anonymous referee for their careful reading of the manuscript and constructive comments. This work was supported by an Australian Research Council Project (ARC DP220102201).

\section*{Data Availability}

The reference catalog and the CNN-based solar flare catalog presented in this work are publicly accessible at \href{https://solarflarecnncatalog.com/}{solarflarecnncatalog.com}. Both catalogs will be archived in the VizieR database to ensure long-term availability for the community.

\clearpage

\appendix

\section{Flare Catalogs and Detection Algorithms} \label{app3}
Solar flare analysis and prediction require algorithms to identify and characterize these events. In the following, we briefly introduce some of the popular flare detection algorithms (FDAs) in solar and stellar studies. Throughout this paper, comparisons are made between the output of some of the FDAs in this Appendix and the CNN results in this paper. However, a detailed quantitative study is deferred to future work, as it is unclear at the time of writing how to make systematic, quantitative, statistically meaningful comparisons when the FDAs are so heterogeneous in their algorithms, detection criteria, confidence metrics, and observing time spans.

\subsection{NOAA FDA and GOES Catalog} \label{A1}
The National Oceanic and Atmospheric Administration (NOAA) algorithm is used to construct the Geostationary Operational Environmental Satellites (GOES) flare catalog \citep{GOES_R_XRS_L2_Products_Guide}. The NOAA GOES FDA ingests the one-minute averaged GOES SXR lightcurves in the long channel which exhibits less noise, and a better signal-to-noise ratio than the short channel \citep{Aschwanden_2012, Janssens2025}. The FDA identifies the start of an event, when four consecutive flux values meet three specific yet arbitrary criteria: they must be above a background threshold, they must be strictly increasing, and the last value must be greater than 1.4 times the first value. The end of the flare is defined to be when the flux decreases to half of its peak value, with the peak time being when the flux reaches its maximum. 

The GOES flare catalog maintains a comprehensive record of solar flares since 1975. The catalog categorizes events based on the flux intensity measured in the 1--8 \AA~ band into A, B, C, M, and X classes. The classification is defined by the peak flux as follows: A-class flares have a flux below $10^{-7}$ W/m\textsuperscript{2}, B-class flares range from $10^{-7}$ to $< 10^{-6}$ W/m\textsuperscript{2}, C-class flares range from $10^{-6}$ to $< 10^{-5}$ W/m\textsuperscript{2}, M-class flares range from $10^{-5}$ to $< 10^{-4}$ W/m\textsuperscript{2}, and X-class flares are those with fluxes greater than or equal to $10^{-4}$ W/m\textsuperscript{2}. The GOES catalog includes information on the start, peak, and end times, as well as the flux values of each flare. Although the original GOES data does not resolve flares spatially, subsequent analysis sometimes includes the approximate location of the flares on the solar disk, typically derived from observations by other instruments such as Solar Dynamics Observatory (SDO). The catalog also links flares to coronal mass ejections (CMEs) and radio bursts. 

Despite its widespread use, the NOAA GOES FDA exhibits certain shortcomings. It often misidentifies the start and end times of events and fails to detect flares occurring in quick succession. Additionally, it fails to locate small events, during periods of high activity with elevated fluxes. Without background subtraction, the peak flux measurements incorporate the total SXR flux from all ARs on the solar disk, which leads to inflated flux values during solar maximum, when multiple ARs exist at any moment. The GOES flare catalog has been a crucial point of reference for studying solar activity and remains a valuable tool for solar physicists and space weather forecasters.

\subsection{RHESSI FDA} \label{A2}
Alongside the GOES mission, the Reuven Ramaty High Energy Solar Spectroscopic Imager (RHESSI) mission \citep{lin2003reuven} observed the Sun in the 6--12 keV energy range and maintains an associated flare catalog from 2002 to 2018. RHESSI’s primary field of view encompassed the entire solar disk. However, it was designed to detect high-energy phenomena regardless of their source, meaning that not all events RHESSI recorded were necessarily of solar origin. Some observations could have been high-energy phenomena from other astrophysical sources or simply instrumental artifacts \citep{Smith2003}. 

The FDA used by RHESSI operates through a series of steps. First, it measures the background count rate using a 60-second running average to set a baseline for detecting flares. If the ratio of the count rate in the front detectors to the total count rate exceeds $3\sigma$ above the background level, the event is considered a flare candidate. This criterion helps to distinguish flare-related emissions from background noise. When data gaps are present, an event is considered valid only if the ratio of the count rate in the front detectors to the total count rate (front-to-total ratio) exceeds 60\%. 

The flare position is determined by reconstructing X-ray maps for various RHESSI’s subcollimators in the 6--12 keV band. For each map, the brightest pixel outside one full-width-at-half-maximum (FWHM) of the instrument’s rotation center, that is, the geometric center of imaging field, is first identified. A provisional centroid is then determined as the median position of these pixels. The flare position is obtained by averaging only those pixels that lie within one FWHM of this centroid, provided that at least three detectors produce mutually consistent results. Finally, a candidate is confirmed as a solar flare if a valid position is found, ensuring the event has occurred on the Sun \citep{nasa_rhessi}.

\subsection{Hinode FDA} \label{A3}
The Hinode Flare Catalog provides a record of solar flares observed by Hinode's Solar Optical Telescope (SOT), X-ray Telescope (XRT), and EUV Imaging Spectrometer (EIS) \citep{kosugi2007hinode}. From September 2006 to late 2011, the Hinode catalog documents approximately 5000 solar flares. Each observed event is then cross-referenced with the RHESSI and Nobeyama Radioheliograph catalogs to ensure accuracy and completeness \citep{watanabe2012hinode}.

\subsection{Aschwanden \& Freeland FDA} \label{A4}
\cite{Aschwanden_2012} introduced a new FDA for SXR flares that offers significantly greater sensitivity than the NOAA GOES FDA. Their method employs the GOES SXR long channel. However, to facilitate a comparison between different observational instruments, SXR data are binned to align with the 12-second cadence of the SDO Atmospheric Imaging Assembly (AIA)  images. This FDA sets a minimum flare duration of one minute to ensure that transient fluctuations are not misclassified as flares and calculates the mean and standard deviation of typical SXR fluxes during quiet periods to establish a reliable noise threshold. The minimum fluxes of daily lightcurves are then subtracted from the signal. This step refines the data by removing daily variations in baseline flux levels. The algorithm also identifies and removes spikes by checking for intervals, where the ratio of the maximum to minimum flux within a 60-second window exceeds a specified threshold. This is necessary to eliminate short-lived anomalies, i.e., spikes, that could distort the flare detection. Further smoothing routines are applied to the lightcurves to remove quasi-periodic pulsations. Finally, the local maxima and minima of the profiles are identified as potential flare start, peak, and end times. Despite the enhanced sensitivity  of this algorithm in the detection of smaller events (up to five times greater than that of the NOAA GOES FDA), it does not distinguish between overlapping flare intervals and consecutive events, because it relies on the slow-driving assumption, whereby one flare must terminate before the next begins \citep{Aschwanden_2012, lu2024automatic, valluvan2024solar}

\cite{Plutino_2023} recently introduced another FDA, which follows a similar approach to the Aschwanden-Freeland method but differs in the sequence of pre-identification steps and the calculation of background flux over 24-hour intervals. It also applies corrections of the SWPC scaling factors for various GOES instruments to obtain true flux values. Plutino’s catalog claims to successfully identify events with overlapping times, but it requires manual decision-making for event classification based on local extrema.

\subsection{TEBBS FDA} \label{A5}
\cite{Ryan_2012} developed a Temperature and Emission Measure-Based Background Subtraction (TEBBS) technique, which isolates the flare-associated flux from the total GOES SXR flux by subtracting the background. The validity of the background choice is rigorously tested through a series of checks. These include temperature and emission measure (EM) tests to confirm that these parameters increase during the rise episode, as expected in typical flare behavior \citep{bornmann1990}, and a hot flare test to ensure that the background-subtracted flux temperature always exceeds the background temperature, remaining within the instrument's operational range (4 MK to 100 MK). An additional pre-processing routine excludes flare-like signals that do not conform to the typical flare behavior, potentially caused by instrumental errors. Therefore, any temperature or EM values from the early rise episode that deviate from expected flare behavior are removed from the dataset. After the pre-processing phases, the TEBBS FDA follows the NOAA GOES steps but benefits from the enhanced accuracy provided by the background subtraction. 

\subsection{SphinX FDA} \label{A6}
Building upon the NOAA GOES and TEBBS FDAs, \cite{gryciuk2017flare} developed a refined FDA utilizing data from the Solar Photometer in X-ray (SphinX) onboard the Complex Orbital Observations Near-Earth of Activity of the Sun-Photon (CORONAS-Photon) spacecraft \citep{kotov2011scientific}. The highly sensitive SphinX instrument, operating in the energy range of 1.2--15 keV (0.8–10~\AA), detects solar events significantly below the detection threshold of the GOES system \citep{sylwester2008sphinx, sylwester2012sphinx}.

Focusing on the period from February to November 2009, during a deep solar minimum, the SphinX FDA employs a semi-automatic process to identify the start, peak, and end times of flares. It begins by smoothing the raw X-ray lightcurves using a boxcar averaging technique over a 70-second interval. The algorithm then identifies peak times by scanning for consecutive increases in flux values. Specifically, it searches for a sequence of four monotonically increasing data points, where the flux at the fourth point exceeds the initial flux by at least 3\%, marking this as the start of an event. Then, the algorithm looks for three consecutive points with decreasing flux values to mark the end of the event. The peak time is associated with the maximum flux occurring between the identified start and end times. A forward and reverse search methodology is used to ensure complete detection.

To determine the start and end times, an Elementary Flare Profile (EFP) model is employed. Accordingly, flares are modeled as a convolution of a Gaussian function (reflecting the energy build-up and release) and an exponential decay function (representing energy dissipation). For consecutive flares, the EFP method decomposes complex events by fitting multiple EFPs to isolate individual components. Additionally, the SphinX FDA incorporates a background subtraction method to account for time-dependent variations. Once the EFP is fitted, the flare profile minus the background is compared to the background's 1-$\sigma$ level. Therefore, the start (end) time is marked where the EFP curve rises above (falls below) the background 1-$\sigma$ level \citep[see Figure 4 in][]{gryciuk2017flare}.  

The SphinX FDA identified 1600 flare events, during its period of study, contrasting significantly with only 9 GOES-detected flares during the same period. After the semi-automatic detection, a manual review was conducted to refine the results. During this step, approximately 500 false detections were removed, and 300 additional flares were identified through visual inspection, yielding a final catalog of 1604 detected events.

\subsection{Smoothing-based FDA} \label{A7}
The FDA introduced by \cite{Sadykov_2019} uses the TEBBS approach for background subtraction and the Savitzky-Golay filter for smoothing GOES SXR lightcurves. It fits a low-degree polynomial to successive subsets of data points, resulting in a smoothed signal. The algorithm then follows the NOAA GOES FDA steps for flare identification.

\subsection{SolO FDA} \label{A8}
SolO FDA detects solar flares in X-ray using data from the Spectrometer/Telescope for Imaging X-rays (STIX) aboard the European Space Agency’s (ESA) Solar Orbiter mission \citep{krucker2020spectrometer, muller2020solar}. Solar Orbiter was launched on February 2020. The first public STIX data were released on 30 September 2020 through the STIX Data Center and ESA’s Solar Orbiter catalog (SOAR). STIX provides full disk HXR imaging spectroscopy over 4–150 keV, offering continuous monitoring and high cadence measurements in up to 32 discrete energy bands.

The SolO FDA works mainly with Quick Look (QL) STIX data. These are low-telemetry lightcurves with 4s cadence and five broad energy bands (4–10, 10–15, 15–25, 25–50, and 50–84 keV). QL data is typically available within minutes to two days of acquisition and includes background estimates, variances, and automated flare flags. The science data, which arrive less frequently (usually within several days), provide finer energy resolution, sub-second cadence, and pixel level detector information for post validation and refinement.

To isolate transient flare emissions from the background, SolO FDA applies the Statistics Sensitive Nonlinear Iterative Peak clipping (SNIP) method \citep{ryan1988snip}. SNIP iteratively compresses and clips the logarithm of peaks to identify a stable background trend, then reverses the transform to recover photon counts. A flare is recorded when the background-subtracted count rate exceeds twice the quiet Sun standard deviation. Start and end times are marked by threshold crossings. Nearby peaks (within 5 minutes time) are merged, and detections are cross-validated across the five energy bands. For flares not seen by GOES, the list also stores an estimated GOES class derived from a cross-calibration with STIX 4–10 keV background-subtracted count rates. The estimation follows a power-law relation between the GOES flux $f$ (in Wm$^{-2}$) and the STIX count rate scaled to 1 AU, $X^{\prime} = x r^{2}$, where $x$ is the background-subtracted count rate, and $r$ is the distance between the Sun and Solar Orbiter (in AU). The corresponding GOES flux is then estimated via $f = 10^{-7.376+0.622\log_{10}(X^{\prime})}$ \citep{xiao2023data}.

\subsection{Fermi-LAT FDA} \label{A9}
The First Fermi-Large Area Telescope (LAT) Solar Flare Catalog details the detection and analysis of 45 solar flares observed by the Fermi-LAT between 2010 to 2018 \citep{Ajello_2021}. The catalog employs the Fermi-LAT SunMonitor automated pipeline, which is designed to monitor the Sun and identify gamma-ray emissions with energies between 30 MeV and 10 GeV. This pipeline operates when the Sun is within 70 degrees of the LAT boresight, maximizing observational coverage.

The Fermi-LAT FDA starts with a maximum likelihood analysis to estimate the significance of gamma-ray detections. The analysis uses a test statistic to confirm a flare, defined as twice the increase in the logarithm of the likelihood when the data are fit with both the source and background components compared to a fit with the background alone \citep{mattox1996likelihood}. For spectral analysis, the gamma-ray data are modeled as a function of energy using a simple power law ($dN/dE \propto E^{-\Gamma}$), a power law with an exponential cutoff ($dN/dE \propto E^{-\Gamma}\exp(-E/E_c)$), and pion-decay emission templates (which describe $\gamma$-rays produced by interactions of accelerated protons with ambient material). These models help differentiate between emissions from various mechanisms, such as bremsstrahlung emissions from electrons and pion-decay emissions from protons. Localization of the flares is achieved by placing a test point source on a grid of sky positions and computing the likelihood ratio at each location. These maps provide a detailed spatial representation of the gamma-ray sources. The temporal analysis involves constructing lightcurves to estimate the start and stop times of gamma-ray emissions, integrating over these periods to calculate total flux and the number of accelerated protons. Finally, the FDA classifies the events into two categories: prompt (coincident with hard X-rays) and delayed (associated with CMEs). 

\subsection{C-PhLARE FDA} \label{A10}
The Colorado Physics Laboratory Academic Research Effort Collaboration (C-PhLARE) has recently developed a systematic method for the identification and analysis of solar flares using GOES data \citep{Mason_2019, mason2023coronal}. The C-PhLARE FDA focuses on defining and subtracting the pre-flare background flux and comprises three main steps. Initially, a pre-flare time window is selected to establish the baseline SXR flux prior to the flare. This window is subsequently divided into three equal sub-windows, with each sub-window being analyzed independently. The median and standard deviation of the flux are calculated within each sub-window. In the absence of variability, the light curve would remain constant throughout the window, resulting in identical median and standard deviation values across all sub-windows. However, in practice, lightcurves often show various trends. Certain conditions are then applied to validate the background level. Provided these conditions are satisfied, the background flux to be removed from each lightcurve is determined as the mean of the three earlier calculated medians. Events that do not meet these conditions are excluded, leaving roughly 30\% of the observations. The impact of different pre-flare time window choices on the number of detected events was also investigated.

\subsection{AGILE FDA} \label{A11}
The First AGILE Solar Flare Catalog provides a record of solar flares detected by the AGILE satellite from May 1, 2007, to August 31, 2022 \citep{Ursi_2023}. This catalog encompasses observations in the 80–200 keV energy range, made by the onboard anticoincidence system, which provides continuous X-ray monitoring along the satellite's orbit. The AGILE FDA involves several steps to identify and characterize solar flares. It begins by collecting high-energy X-ray transients at 1.024-second intervals. Transients are identified as potential solar flares if their signal exceeds a threshold of $3\sigma$ above the background rate for at least 60 consecutive frames (approximately one minute). This criterion helps exclude short-duration spikes caused by high-energy particles or other transient phenomena, such as gamma-ray bursts or soft gamma-ray repeaters. The first frame that exceeds the threshold is marked as the start of the flare, and the last frame is marked as the end. The highest count rate between these start and end times is identified as the peak. To confirm the solar origin of the detected transients, each event is cross-referenced with the GOES solar flare database. A match is confirmed if the maximum of the AGILE transient falls within the start and end times of a GOES-listed flare. In cases where the AGILE event is saturated, any part of the saturated segment that falls within the GOES interval is considered a match. Matched flares are then classified based on the GOES catalog. The catalog also checks against other high-energy missions, such as RHESSI and Fermi GBM. Notably, this catalog includes information on 1424 events that are not registered in the GOES database, likely representing low-intensity, short-duration flares.

\subsection{Radio Frequency FDA} \label{A12}
Another FDA based on perturbations in very low frequency (VLF) radio flux was introduced by \cite{George_2019} and further developed by \cite{Belcher_2021}. This method uses ground-based VLF radio measurements as proxies for solar X-ray data and identifies significant M and X-class flares by analyzing ionospheric observations. During quiet periods, the solar X-ray flux is too small to ionize the D region (the lowest part of the Earth's ionosphere, approximately 60 to 95 kilometers above the Earth's surface). This flux, along with other galactic sources, only ionizes nitric oxide. However, during solar flares, additional ionospheric constituents, including N\textsubscript{2} and O\textsubscript{2}, become ionized. The increased ionization reduces the effective reflection height for VLF waves, causing perturbations in both VLF amplitude and phase. Modeling the solar X-ray flux by VLF perturbations enables the rapid detection and classification of elevated X-ray levels. This FDA effectively identifies large flares occurring during daytime when the solar zenith angle (SZA) is less than 85 degrees. The SZA constraint avoids sunrise and sunset conditions, which produce large-scale ionization changes. No saturation was observed in the response of VLF phase and amplitude to increasing solar X-ray flux, demonstrating that VLF phase perturbations are effective for detecting intense solar flares \citep{thomson2005large}.

\subsection{Heliophysics Event Knowledgebase FDA} \label{A13}
Heliophysics Event Knowledgebase (HEK) serves as another catalog for solar flares, using data from the Solar Dynamics Observatory (SDO) \citep{Guide_SDO_Data_Analysis, Martens_2012}. The HEK FDA is designed to provide near real-time flare alerts and generate detailed statistical surveys of flaring events. It primarily uses Atmospheric Imaging Assembly (AIA) full-disk images in the 193 \AA~ and segments each image into 16 × 16 macropixels. At times, hotter bands, including 94 and 131 \AA~ are used as well. A peak-detection algorithm, adapted from the RHESSI FDA, is then applied to the integrated signal of each macropixel. This involves smoothing the lightcurves and detecting local maxima. Thus, flare parameters such as start, peak, and end times, location (when available), associated AR NOAA number (when available), and peak intensity are extracted for each macropixel. Plasma parameters, including temperature and EM, are calculated using the background-subtracted data. To validate the background subtraction accuracy, the HEK FDA performs several tests on the temperature and EM similar to the TEBBS algorithm.

The HEK flare catalog also records information on event classes, based on the GOES classification system (if applicable), and links flares with other solar phenomena like CMEs and radio bursts. The HEK FDA detects flares as small as C1-class and supports the identification of simultaneous flares in different ARs.

\subsection{Alternative EUV-driven FDA} \label{A14}
An alternative EUV-based FDA was developed by \cite{Van_der_Sande_2022}, which detects solar flares using SDO/AIA cutout images across six channels: 94, 131, 171, 193, 304, and 1600 \AA. Each image is reduced to a scalar number representing the sum of flux over the disk, resulting in six scalar time series that peak during flaring events. These time series are then resampled to a one-minute cadence, except for the 1600 \AA~ channel, which has a 72-second cadence. Peaks of the flaring events are identified as the highest peak in a time window, with start times determined from the first and second derivatives of the flux. The end times are also marked when the intensity drops by 80\% relative to the peak value. This FDA compares flare characteristics across channels, requiring overlapping peaks in at least three of four wavelengths. This catalog shows an overlap of 85\% with the GOES flare catalog for M- and X-class events.

\subsection{Other FDAs}
Flare forecasting relies on detecting and understanding flare events to predict their likelihood. Forecasting methodologies are often built on pre-flare characteristics, such as sunspot number \citep{giovanelli1939}, magnetic field gradient \citep{cui2006}, neutral line length \citep{falconer2001}, energy dissipation \citep{song2009}, or total and free magnetic energy and helicity \citep{thalmann2025}. These methodologies can be broadly categorized as FDAs. Machine learning techniques, including support vector machines \citep{nishizuka2017}, random forests \citep{florios2018}, Long Short-term Memory (LSTM) \citep{hassani2025solar}, and CNNs \citep{huang2018}, have also enabled precursor detection in photospheric magnetograms, EUV images, and SXR lightcurves \citep{krista2021, sun2023, krista2024}.

Furthermore, numerous methodologies have been proposed for detecting stellar flares using data from missions such as the Transiting Exoplanet Survey Satellite (TESS). One approach is that of \cite{Crowley_2022}, which segments the lightcurve into intervals, applies sigma-clipping, normalizes by median flux, fits a third-order polynomial for long-term trends, and then uses the Lomb-Scargle periodogram to exclude periodic modulations, followed by a nonlinear filter to detect transient increases indicative of flares. Other approaches are also possible such as detrending with spline fits and detecting flares based on $\sigma$ thresholds \citep[see e.g.,][]{Gunther_2020}, as well as CNNs to identify flares without the need for detrending \citep[see e.g.,][]{Feinstein_2020}.

\section{CNN Implementation Details} \label{app4}
The CNN architecture integrates convolutional, recurrent, and attention-based modules to capture both local and global temporal patterns in the GOES SXR flux signal \citep{goodfellow2016deep, vaswani2017attention, bai2018empirical, el2025hybrid}. Input windows, consisting of fixed-length normalized flux time series, are processed in parallel through four convolutional branches with kernel sizes of 3, 5, 7, and 11. Each branch comprises four convolutional layers with channel depths of 16, 32, 64, and 128, with rectified linear unit (ReLU) activations applied after each layer. Unlike conventional CNN pipelines, no pooling or downsampling operations are applied. This design choice is motivated by the need to preserve precise flare start and peak times. To prevent overfitting, a single dropout layer with a rate of 0.2 is introduced after the convolutional blocks.

The resulting feature maps, which are the summaries of patterns captured by convolutional filters, are then modeled sequentially by a bidirectional long short-term memory (BiLSTM) layer with 256 hidden units in each direction, which captures gradual flux changes and short-range dependencies by incorporating information from both past and future time steps \citep{huang2015bidirectional, siami2019performance}. The BiLSTM was initially introduced to enforce temporal ordering constraints in the flare evolution, because the network design we tried first aims to identify the complete flare profile comprising both rise and decay episodes and background, i.e., three prediction classes. Although the final model adopts a simplified two-class formulation (rise versus background), the BiLSTM is retained, as architectures incorporating this component shows better performance than basic configurations during model development.

To complement this, the sequence is further processed by a Transformer encoder, whose self-attention mechanism captures long-range relations \citep{vaswani2017attention, han2021transformer}. The encoder consists of a single layer with a 512-dimensional embedding space, eight attention heads, and a feed-forward dimension of 1024. Standard sinusoidal positional encoding is used to provide temporal context prior to attention. 

Although the BiLSTM and Transformer are often used independently, their combination here is deliberate: the BiLSTM provides local sequence modeling, while the Transformer supplies global context, and together they achieve better performance in our experiments than either module alone \citep{khan2024transformer, xiong2024load, WANG2025137654}. The final classification is performed by a fully connected layer \citep{ma2017equivalence}. 

The CNN architecture is implemented in PyTorch and trained using a weighted cross-entropy loss to address class imbalance between flaring and nonflaring samples \citep{phan2020resolving}. Optimization is performed using the Adam algorithm, a stochastic gradient–based optimization method that adapts individual learning rates for each parameter, with a learning rate of $10^{-5}$ and a mini-batch size of 16.

\bibliography{refs.bib} 

% \bibliography{sample701}{}
\bibliographystyle{aasjournalv7}

%% This command is needed to show the entire author+affiliation list when
%% the collaboration and author truncation commands are used.  It has to
%% go at the end of the manuscript.
%\allauthors

%% Include this line if you are using the \added, \replaced, \deleted
%% commands to see a summary list of all changes at the end of the article.
%\listofchanges

\end{document}